\documentclass[aps,prd,nofootinbib,twocolumn,groupedaddress]{revtex4}
\usepackage{dcolumn}
\usepackage{bm}
\usepackage{pdfpages}
\usepackage{amsmath}    % need for subequations
\usepackage{amsfonts} % note how statements can be commented out
\usepackage{amssymb}
\usepackage{mathptmx}      % use Times fonts if available on your TeX system
\usepackage{graphicx}   % for figures
\usepackage{mathtools}
\usepackage{graphics}
\usepackage{epsfig}
\usepackage{tabularx}   % fit table to page width
\usepackage{multirow}
\usepackage{bibentry}
\usepackage{braket}
\usepackage{algorithm}
\usepackage{algorithmic}
\usepackage{listings}
\usepackage{setspace}
\usepackage{fancyhdr}
\usepackage{slashed}
\usepackage{color}
\usepackage{xfrac}
\usepackage[pdftex, pdfborder= 0 0 0, citecolor=blue, urlcolor=blue, linkcolor=blue, colorlinks=true, bookmarksopen=true]{hyperref}
\def\Journal#1#2#3#4{{#1} {\bf#2}, #3 (#4)}

\def\NPA{{\rm Nucl. Phys.} A}
\def\NPB{{\rm Nucl. Phys.} B}
\def\PLB{{\rm Phys. Lett.}  B}
\def\PRL{\rm Phys. Rev. Lett.}
\def\PRD{{\rm Phys. Rev.} D}
\def\PRC{{\rm Phys. Rev.} C}

\def\JPG{{\rm J. Phys.} G}
\def\EPJC{{\rm Eur. Phys. J.} C}
\def\EPJA{{\rm Eur. Phys. J.} A}

\def\IJMPA{{\rm Int. J. Mod. Phys.} A}
\def\ep{\epsilon}
\def\vep{\varepsilon}
\def\la{\langle}
\def\ra{\rangle}

\def\lam{\lambda}
\def\al{\alpha}

\def\be{\begin{equation}}
\def\ee{\end{equation}}
\def\bea{\begin{eqnarray}}
\def\eea{\end{eqnarray}}
\begin{document}

\title{Spacelike and timelike form factors for the $(\pi^0,\eta,\eta')\to\gamma^*\gamma$ transitions in the light-front quark model}
\author{ Ho-Meoyng Choi\\
{\em Department of Physics, Teachers College, Kyungpook National University,
     Daegu 41566, Korea}\\
     Hui-Young Ryu\\
{\em Department of Physics, Pusan National University,
     Pusan 46241, Korea}\\         Chueng-Ryong Ji\\
{\em Department of Physics, North Carolina State University,
Raleigh, North Carolina 27695-8202, USA} }
\begin{abstract}
We investigate the $(\pi^0,\eta,\eta')\to\gamma^*\gamma$ transitions both for the spacelike region and the timelike region using the 
light-front quark model (LFQM). In particular, we present the new direct method to explore the timelike region without resorting to mere analytic continuation
from the spacelike region to the timelike region. Our direct calculation in timelike region shows the complete agreement not only with 
the analytic continuation result from the spacelike region but also with the result from the dispersion relation between
the real and imaginary parts of the form factor.
For the low energy regime, we compare our LFQM results of the transition form factors (TFFs) for the low timelike momentum 
transfer region and the slope parameters at $q^2=0$ with the recent experimental data from the Dalitz decays of $(\pi^0,\eta,\eta')$.
For the high energy regime, we incorporate the QCD factorization in our LFQM to examine the asymptotic behavior of TFFs 
both for the spacelike region and the timelike region. We compare our results with the available experimental data. 

\end{abstract}
\maketitle

\section{Introduction}
The meson-photon transition form factors (TFFs) such as $F_{P\gamma}(Q^2)(P=\pi^0,\eta,\eta')$ have been known to be
the simplest exclusive processes involving the strong interaction. 
They play a significant role in allowing both the low- and high-energy precision tests of 
%%HM: remove (SM)
the standard model, 
%%%
in particular,
the quantum chromodynamics (QCD)\cite{Czerwinski}.

For the low-energy regime, the  TFFs  enter the prediction of important observables such
as the rates of rare decays $P\to\bar{\ell}\ell(\ell=e,\mu)$~\cite{HL15} and
the hadronic light-by-light (HLbL) scattering contribution to the muon anomalous magnetic moment $(g-2)_\mu$~\cite{JN,Ny2016}. 
The HLbL contribution is in principle obtained by integrating some weighting functions times the product of a single-virtual and
a double-virtual TFFs for  spacelike momentum~\cite{JN,Lattice16}. 
While there are currently no available data for the double-virtual TFFs,
the single-virtual TFFs are available from the $\gamma^*\gamma\to(\pi^0,\eta,\eta')$ processes in the small and intermediate momentum 
transfer range up to $Q^2\sim 8$ GeV$^2$. The $(\pi^0,\eta,\eta')\to\gamma^*\gamma$ TFFs 
of the spacelike regon have been measured experimentally by several collaborations~\cite{CELLO91,CLEO98,BES15_Pi}.
Recently, the single-virtual TFFs 
for small timelike momentum transfer ($q^2=-Q^2>0$) regions and the slope parameters at $q^2=0$
have also been measured~\cite{NA60,NA60-17,A22014, A22011,A2pi,BES15} from the Dalitz decays $P\to\bar{\ell}\ell\gamma$ 
where $(2m_\ell)^2\leq q^2\leq m^2_P$.

For the high-energy regime, the TFFs can be calculated asymptotically at leading twist as a convolution of the perturbative 
hard scattering amplitude and the gauge-invariant meson distribution 
amplitude (DA)~\cite{BL80,ER80,CZ84} which incorporates the nonperturbative dynamics of QCD bound state.
In particular, hadronic DA~\cite{BL80,ER80,CZ84}  provides an essential
information on the QCD interaction of quarks, antiquarks and gluons inside the hadrons and
plays an essential role in applying QCD to hard exclusive processes.
The prediction for the single-virtual pion TFF, $F_{\pi\gamma}(Q^2)$, at the asymptotic limit $Q^2\to\infty$ is 
given by the well-known Brodsky-Lepage limit~\cite{BL80}:
$Q^2 F_{\pi\gamma}(Q^2\to\infty)=\sqrt{2}f_{\pi}\simeq 0.185$ GeV.
However, the BaBar Collaboration~\cite{Babar09} has measured the $F_{\pi\gamma}(Q^2)$ up to about $Q^2\sim 35$ GeV$^2$
from reaction $e^+e^-\to e^+e^-\pi^0$ in the single tag mode and 
have shown not only the serious violation of the Brodsky-Lepage limit but also the rapid growth for $Q^2>15$ GeV$^2$.
On the other hand, the subsequent Belle Collaboration~\cite{Belle12} has reported their measurement
for $F_{\pi\gamma}(Q^2)$ and has shown that the measured values of $Q^2F_{\pi\gamma}(Q^2)$
are consistent with the asymptotic limit of QCD for $Q^2>15$ GeV$^2$.
For the reaction $e^+e^-\to e^+e^-\eta^{(\prime)}$,  the subsequent BaBar
data~\cite{Babar11} for the $Q^2F_{\eta^{(\prime)}\gamma}(Q^2)$ TFFs 
provided a consistency with the perturbative QCD prediction unlike the case of pion TFF~\cite{Babar09}.
These discrepancies for the results of $Q^2 F_{\pi\gamma}(Q^2)$  between the BaBar and
the Belle data  and between $Q^2 F_{\pi\gamma}(Q^2)$  and $Q^2F_{\eta^{(\prime)}\gamma}(Q^2)$ TFFs for $Q^2>15$ GeV$^2$
region have motivated many theoretical studies using various forms of the meson DAs to
understand and reconcile those discrepancies~\cite{MS09,Ra09,MP09,AB03,Dor10,DK13,WH10, Kroll11,ABOP,CD10,DSE16,BCT,BMPS11,SBMP,LM12,BLM12,SSK,KV10,LSW,Hwang01,LG12,MBF13,MBF,DNSS,NV10,FK98,DRZ,AS04,eta1,eta2,eta3,eta4,Wang17}.

To examine the issue of the scaling behavior of $Q^2 F_{\pi\gamma}(Q^2)$ in the large $Q^2$, it may be necessary to analyze the corresponding form factor not only in the spacelike region but also in the timelike region.
To explore the timelike region beyond the single Dalitz decays ~\cite{NA60,NA60-17,A22014, A22011,A2pi,BES15}, 
the $e^+e^-$ colliders access the values $q^2 > m^2_P$ through the $e^+e^- \to P\gamma$ annihilation processes.
Although the data for $F_{\pi\gamma}(q^2)$ in the large timelike $q^2$ region is not available yet,
the BaBar Collaboration~\cite{BaBar06} measured the timelike $F_{\eta^{(\prime)}\gamma}$ TFFs
from the reaction $e^+e^-\to\eta^{(\prime)}\gamma$ at an average $e^+e^-$ center of mass energy of
$\sqrt{s}=10.58$ GeV, which corresponds to $q^2=112$ GeV$^2$.
However, the theoretical analysis for the timelike region going beyond $q^2 > m^2_P$ is highly nontrivial 
due to the singular nature and the complexity of the timelike form factor. 
Some theoretical subtleties regarding on the analytic continuation 
from the spacelike region to the timelike region can be found in~\cite{Kroll11,BRS00}.
While some theoretical analyses~\cite{EMSeta,EMSetap} for the TFFs in timelike region can also be found for some Dalitz 
decays ($(2m_\ell)^2\leq q^2\leq m^2_P$), we do not yet find any theoretical analysis going beyond 
the Dalitz decay region, i.e. $q^2 > m^2_P$ region. 

Thus, we attempt to explore the entire timelike region as well as the spacelike region in this work.
We extend our previous analysis~\cite{CJ_99,CJ_DA,PiGam16}
for the single-virtual $P\to\gamma^*\gamma ~(P=\pi^0,\eta,\eta')$ transition in the spacelike 
region using the light-front quark model (LFQM)~\cite{CJ_99,CJ_DA,PiGam16,CJ_PLB,CJBc} to include the entire timelike
region. For the low energy regime, we compare our LFQM results of the TFFs for the low timelike momentum transfer region and 
the slope parameters at $q^2=0$ with the recent experimental data from the
%%HMC  
Dalitz 
decays of $(\pi^0,\eta,\eta')$~\cite{NA60,NA60-17,A22014, A22011,A2pi,BES15}.
For the high energy regime, we show the asymptotic behavior of TFFs for both space- and time-like regions and compare them with the
available experimental data. 
In particular, we present the new direct method to explore the timelike region without resorting to mere analytic continuation
from space- to time-like region. Our direct calculation in timelike region shows the complete agreement with 
not only the analytic continuation result from spacelike region but also the result from the dispersion relation (DR) between
the real and imaginary parts of the form factor.

The paper is organized as follows. In Sec.~\ref{sec:II}, we discuss the
meson-photon TFFs in an exactly solvable model first based on the covariant 
%%HMC
Bethe-Salpeter~(BS)
%%%
model of
(3+1)-dimensional fermion field theory. It has been a common practice to utilize an exactly solvable
manifestly 
%%HMC
covariant BS model 
%%%
to check the existence (or
absence) of the LF zero mode~\cite{Zero1,Zero2,Zero3,Zero4}
as one can pin down the zero mode exactly in the manifestly covariant 
BS model~\cite{BCJ02,BCJ03,TWV, TWPS,TWPS17}. 
Performing both manifestly covariant
calculation and the LF calculation, we explicitly show the equivalence between the
two results and the absence of the zero-mode contribution to the TTF. 
In the LF calculation, we analyze both $q^+(=q^0+q^3)\neq 0$
and $q^+=0$ frames and show their equivalence in the numerical calculation.
We explicitly demonstrate that our direct LFQM result for the timelike form factor  
is in complete agreement with the result obtained from the DR method.
The $\eta-\eta'$ mixing scheme for the calculations of the $(\eta,\eta')\to\gamma^*\gamma$
TFFs is also introduced in this section.
In Sec.~\ref{sec:III}, we apply the self-consistent correspondence relations~(see, e.g., Eq. (35) in~\cite{TWPS}) between the covariant BS model and the LFQM and present the standard LFQM calculation with 
the more phenomenologically accessible model wave functions 
provided by the LFQM analysis of meson mass spectra~\cite{CJ_PLB,CJ_99}.
The self-consistent covariant descriptions of the meson TFFs 
are confirmed in the standard LFQM as we discuss in this section.
In Sec.~\ref{sec:IV}, we present our numerical results for the
$(\pi^0,\eta,\eta')\to\gamma^*\gamma$ TFFs for both spacelike and timelike regions 
and compare them with the available experimental data. Summary and discussion follow in Sec.~\ref{sec:V}. In the Appendix, we provide the comparison of the
$\eta-\eta'$ mixing angle between the octet-singlet basis and quark-flavor basis.

\section{Manifestly Covariant Model}
%\subsection{Two-point function: decay amplitude}
\label{sec:II}
\begin{figure*}
\begin{center}
\includegraphics[height=3.5cm, width=17cm]{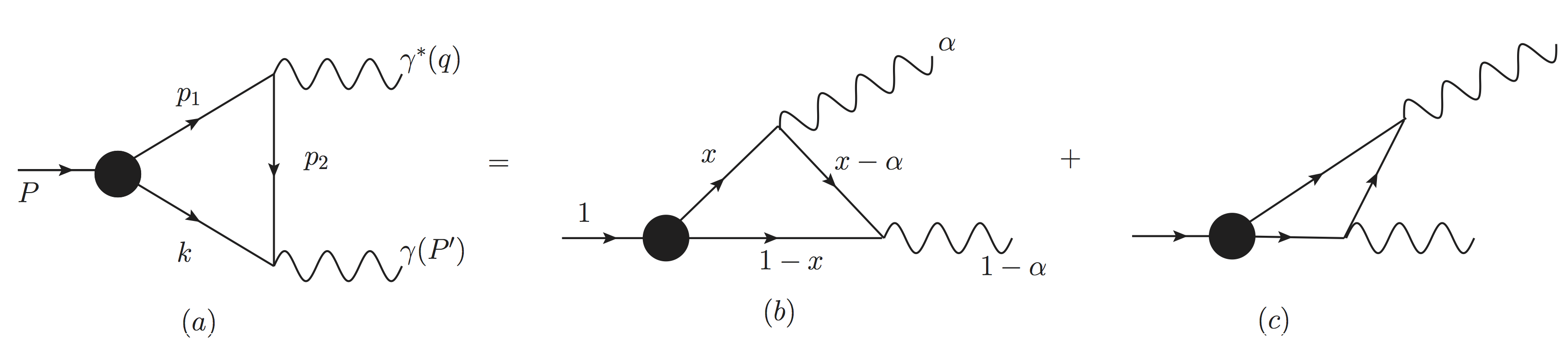}
\caption{\label{fig1} 
{\color{black}
One-loop Feynman diagrams that contribute to $P\to\gamma^*\gamma$. The single covariant Feynman diagram (a) is the same
as the sum of the two LF time-ordered diagrams (b) and (c), respectively. 
}
}
\end{center}
\end{figure*}
The transition form factor $F_{P\gamma}$ for 
%%%HMC: 
the $P\to\gamma^*\gamma$~($P=\pi^0, \eta, \eta'$)
%%%
decay is defined from
the matrix element of electromagnetic current $\Gamma^\mu=\la\gamma(P-q)|J^\mu|P(P)\ra$
as follows:
\be\label{Eq1}
\Gamma^\mu = i e^2 F_{P\gamma}(Q^2)\ep^{\mu\nu\rho\sigma}P_\nu\vep_\rho q_\sigma,
\ee
where $P$ and $q$ are the momenta of the incident pseudoscalar meson and virtual photon,
respectively, and $\vep$ is the transverse polarization vector of the final (on-shell)
photon. This process is illustrated by the Feynman diagram in Fig.~\ref{fig1} (a),
which represents the amplitude of the virtual photon being attached to the
quark line. 
While we shall only discuss the amplitude shown in Fig.~\ref{fig1} (a),
the total amplitude should of course include the contribution from the amplitude of the virtual photon being attached to
the antiquark line as well as the quark line. 

In the exactly solvable manifestly covariant BS model, the covariant amplitude $\Gamma^\mu$ in Fig.~\ref{fig1} (a)
is obtained by the following momentum integral
\be\label{Eq2}
\Gamma^\mu = i e_Q e_{\bar Q} N_c
\int\frac{d^4k}{(2\pi)^4} \frac{H_0}
{N_{p_1} N_k N_{p_2}}S^\mu,
\ee
where $N_c$ is the number of colors and $e_{Q(\bar Q)}$ is the quark~(antiquark) electric charge.
The denominators $N_{p_j} (= p_{j}^2 -m_Q^2 +i\ep) (j=1,2)$
and $N_k(= k^2 - m_{\bar Q}^2 + i\ep)$ come from the intermediate quark and antiquark propagators
of mass $m_Q=m_{\bar Q}$ carrying the internal
four-momenta $p_1=P-k$, $p_2=P-q-k$, and $k$, respectively.
The trace term $S^\mu$ in Eq.~(\ref{Eq2}) is obtained as
\bea\label{Eq3}
 S^\mu &=&  {\rm Tr}\left[\Gamma_P\left(\slash \!\!\!\!\!p_1 + m_Q \right)
 \gamma^\mu \left(\slash \!\!\!\!\!p_2 + m_Q \right)\slash \!\!\!\!\!\ep
 \left(-\slash \!\!\!\!k + m_Q \right) \right]
 \nonumber\\
&=& 4 i m_Q \ep^{\mu\nu\rho\sigma} \{ q_\nu \ep_\rho k_\sigma
 + (P-k)_\nu q_\rho \ep_\sigma \},
\eea
where we use $\Gamma_P=\gamma_5$ for the pseudoscalar vertex structure.
For the ${\bar q}q$ bound-state vertex function $H_0=H_0 (p^2_1, k^2)$ of the
meson, we simply take 
%%%CRJ
the {\color{black} dimensionless} constant parameter $g$ 
%%%CRJ
since the covariant loop
is regularized with this constant vertex in our model calculation.

Using the following Feynman parametrization for the three propagators
 \bea\label{Eq4}
\frac{1}{N_{p_1} N_{k} N_{p_2}} &=&
\int^1_0 dx \int^{1-x}_0  dy
\nonumber\\
&&\times
\frac{2}{[N_k + (N_{p_1}  -  N_k) x + (N_{p_2} - N_k) y]^3},
\nonumber\\
 \eea
and shifting the variable $k$ to $k'= k-(x+y)P + y q$,
we obtain the manifestly covariant result by defining the amplitude in Fig.~\ref{fig1} (a) as
$\Gamma^\mu_{(a)} = ie_Q e_{\bar Q} [I^{m_Q}_{(a)}]^{\rm Cov}(q^2)\ep^{\mu\nu\rho\sigma}P_\nu\vep_\rho q_\sigma$,
where
\be\label{Eq6}
[I^{m_Q}_{(a)}]^{\rm Cov}= \frac{N_c g}{4\pi^2} \int^1_0 dx
\int^{1-x}_0 dy
\frac{m_Q}{x(1-x-y)M^2 + xy q^2 - m^2_Q},
\ee
with the physical meson mass $M$. 
Similarly,  the amplitude  of the photon being attached to the antiquark line is obtained
by changing $x \to 1 - x- y$ in Eq.~(\ref{Eq6}) but the two results are found to give the same numerical values.
Thus, we obtain the total result  as $I^{m_Q}_{\rm tot}= 2 [I^{m_Q}_{(a)}]^{\rm Cov}$.

For the LF calculation in parallel with the manifestly covariant one,
we use  the plus component $(\mu=+)$ of the currents $J^\mu$ but with two different reference frames, i.e., (1) $q^+\neq 0$ frame and (2) $q^+=0$ 
frame.

In the $q^+\neq 0$ frame, we take $P=(P^+, P^-, {\bf P}_\perp)=( P^+, M^2/P^+, 0)$ and
$q=(q^+, q^-, {\bf q}_\perp)=(\alpha P^+, M^2/P^+, 0)$ so that  $q^2 =q^+q^-$ of the virtual photon
is given by
\be\label{q1}
q^2 =\alpha M^2,
\ee
where $\alpha=q^+/P^+=1-P'^+/P^+$. We should note that $q=(\alpha P^+, M^2/P^+, 0)$ and $P'=P-q=((1-\al)P^+,0,0)$ are valid only for $\alpha\neq 1$
but will differ for the $\al\to 1$ limit as we shall discuss shortly.
In this $q^+\neq 0$ frame, the Cauchy integration over $k^-$ in Eq.~(\ref{Eq2}) has two nonzero contributions to the
residue calculations, i.e., one coming from the interval (i) $0 < k^+ < P'^+$ (see Fig.~\ref{fig1} (b)) and the other from
(ii) $P'^+<k^+< P^+$ (see Fig.~\ref{fig1} (c)).
That is, the Feynman covariant diagram in Fig.~\ref{fig1} (a) is equivalent to the sum of
two LF time-ordered diagrams in Figs.~\ref{fig1} (b) and~\ref{fig1} (c). 
The internal momentum $k^+$ is defined by $k^+=(1-x)P^+$, where $x$ is the Lorentz invariant longitudinal 
momentum variable. 
In this case, the four momenta of the on-mass-shell
quark ($p^2_{\rm 1on}=m^2_Q$) and antiquark ($k^2_{\rm on}=m^2_{\bar Q}$) propagators are defined by
$p_{1\rm on}=(x P^+, p^-_{1\rm on}, -{\bf k}_\perp)$
and
$k_{\rm on}=((1-x) P^+, k^-_{\rm on}, {\bf k}_\perp)$,
respectively.

While  the residue is at the pole of $k^- = k^-_{\rm on}$, which is placed 
in the lower half of complex-$k^-$ plane for the region of $0 < k^+ < P'^+$(see Fig.~\ref{fig1} (b)),
the residue is at the pole of $p^-_1 = p^-_{1\rm on}$, which is placed 
in the upper half of complex-$k^-$ plane for the region of  $P'^+<k^+< P^+$ (see Fig.~\ref{fig1} (c)).
Thus, by defining the amplitude 
$[\Gamma^+_{(b,c)}]^{\rm LF} \equiv ie_Q e_{\bar Q} [I^{m_Q}_{(b,c)}]^{\rm LF}_{\al}(q^2)\ep^{+\nu\rho\sigma}P_\nu\vep_\rho q_\sigma$
for Figs.~\ref{fig1} (b)~and~\ref{fig1} (c), the Cauchy integration of Eq.~(\ref{Eq2}) over $k^-$ in the two regions yields
\be\label{NEq7}
[I^{m_Q}_{(b)}]^{\rm LF}_{0<\al<1}=\frac{N_c}{4\pi^3}\int^1_{\alpha} \frac{dx}{(1-x)} \int d^2{\bf k}_\perp \frac{m_Q}{(\alpha-1) M^2_0}\chi(x,{\bf k}_\perp),
\ee
and 
\bea\label{NEq8}
[I^{m_Q}_{(c)}]^{\rm LF}_{0<\al<1}&=&\frac{N_c}{4\pi^3}\int^{\alpha}_0 \frac{x' dx}{(1-x)} 
\nonumber\\
&&\times \int d^2{\bf k}_\perp 
\frac{m_Q\; \chi(x,{\bf k}_\perp)}{ x' (1-x')M^2-x(1-x)M^2_0},
\eea
respectively, where $x'=x/\alpha$ and
\be\label{NEq9}
\chi(x,{\bf k}_\perp) = \frac{g}{x (M^2 -M^2_0)},
\ee
with 
%\label{NEq10}
 $M^2_0 = \frac{ {\bf k}^{2}_\perp + m^2_Q}{x (1-x)}$
%\ee
being the invariant mass\footnote{
For the calculation of the trace term $S^\mu$ with $\mu=+$, since the result is given by 
$S^+=8 i m_Q \ep_{+-xy}P^+ ({\bf\ep}_\perp \times{\bf q}_\perp)$, one should first take ${\bf q}_\perp\neq 0$ 
and then take ${\bf q}_\perp\to 0$ limit at the end of the trace calculation.}.
We confirmed numerically that $[I^{m_Q}_{(a)}]^{\rm Cov}=[I^{m_Q}_{(b)}]^{\rm LF}_{0<\al<1}+[I^{m_Q}_{(c)}]^{\rm LF}_{0<\al<1}$ as expected.

Now, we find very interesting LF result by taking $\alpha\to1$ limit, 
which in fact allows our direct calculation of the timelike TFFs in LFQM possible
as we present in the next section, Sec.~\ref{sec:III}.
In the $\alpha =1$ case, the four momenta
$q$ and $P'$ are given by $q=(P^+, q^2/P^+, 0)$ and $P'=(0, (M^2-q^2)/P^+,0)$, respectively.
Since $q$ and $P'$ in the $\alpha = 1$ case are different from the  $\al\neq 1$ (i.e. $q^2=\alpha M^2$) case, 
one should not directly substitute $\alpha = 1$ in Eqs.~(\ref{NEq7}) and~(\ref{NEq8}) to obtain the transition amplitudes corresponding
to the $\al = 1$ case.
To obtain the amplitude for the $\al = 1$ case, one needs to go back and start from Eq.~(\ref{Eq2}) again
to do the Cauchy integration over $k^-$ with the specifically given four momenta
$q$ and $P'$ at $\al = 1$.
By doing the Cauchy integration of Eq.~(\ref{Eq2}) over $k^-$ in the $\al = 1$ case, we find
$[I^{m_Q}_{(b)}]^{\rm LF}_{\alpha = 1}=0$ and
\be\label{NEq10}
[I^{m_Q}_{(c)}]^{\rm LF}_{\alpha =1}=\frac{N_c}{4\pi^3}\int^1_0 \frac{dx}{(1-x)^2} \int d^2{\bf k}_\perp 
\frac{m_Q}{M^2_0-q^2}\; \chi(x,{\bf k}_\perp).
\ee
That is, we find in the $q^+\neq 0$ frame at $\alpha = 1$ that  only Fig.~\ref{fig1}~(c) contributes to the total transition amplitude.
We also numerically confirm that Eq.~(\ref{NEq10}) exactly coincides with the manifestly covariant result $I^{m_Q}_{(a)}$ given
by Eq.~(\ref{Eq6}) as it must be. 

For the $q^+=0$ frame,  we take $P=( P^+, M^2/P^+, 0)$ and
$q=(0, q^-, {\bf q}_\perp)$ so that  $q^2 =-{\bf q}^2_\perp \equiv -Q^2$.
Since this $q^+=0$ frame essentially corresponds to the $\alpha\to 0$ limit but with ${\bf q}_\perp\neq 0$, we
refer this frame as the $\al = 0$ case in contrast to the $\al =1 $ case discussed above.
In the $\al = 0$ case, we find that
only Fig.~\ref{fig1} (b) contributes and
the Cauchy integration of Eq.~(\ref{Eq2}) over $k^-$ in Fig.~\ref{fig1} (b) yields
\be\label{Eq8}
 [I^{m_Q}_{(b)}]^{\rm LF}_{\al = 0} = \frac{N_c }{4\pi^3}\int^{1}_0
 \frac{dx}{x(1-x)} \int d^2{\bf k}_\perp
 \frac{m_Q}{M^{\prime 2}_0} \chi(x,{\bf k}_\perp),
\ee
where $M'_0=M_0 ({\bf k}_\perp\to {\bf k}'_\perp)$
with ${\bf k'}_\perp = {\bf k}_\perp + (1-x){\bf q}_\perp$.
We again confirmed numerically that Eq.~(\ref{Eq8}) exactly coincides with the manifestly covariant result $I^{m_Q}_{(a)}$ given
by Eq.~(\ref{Eq6}) as it must be. 
Effectively, we obtain
$[I^{m_Q}_{(a)}]^{\rm Cov}=[I^{m_Q}_{(c)}]^{\rm LF}_{\al =1}=[I^{m_Q}_{(b)}]^{\rm LF}_{\al = 0}=[I^{m_Q}_{(b)}]^{\rm LF}_{0<\al<1}+[I^{m_Q}_{(c)}]^{\rm LF}_{0<\al<1}$.
This result verifies also the absence of the LF zero-mode in pseudoscalar meson TFFs,
i.e., $[I^{m_Q}_{(b)}]^{\rm LF}_{\alpha = 1}=0$ and $[I^{m_Q}_{(c)}]^{\rm LF}_{\al = 0}=0$.

For $(\eta,\eta')\to\gamma^*\gamma$ transitions, we take into account the presence of 
the strange quark and antiquark components in the $\eta$ and $\eta'$ mesons as well as their mixing
with the non-strange quark and antiquark components. Making use of the
$\eta-\eta'$ mixing scheme (see Appendix),
the flavor assignment of $\eta$ and $\eta'$ mesons in the quark-flavor basis $\eta_q=(u\bar{u}+d\bar{d})/\sqrt{2}$ and
 $\eta_s=s\bar{s}$ is given by~\cite{FKS,OZI,Leut98,SSW,Jaus91}
  \be\label{Eq7a}
 \left( \begin{array}{cc}
 \eta\\
 \eta'
 \end{array}\,\right)
 =\left( \begin{array}{cc}
 \cos\phi\;\; -\sin\phi\\
 \sin\phi\;\;\;\;\;\cos\phi
 \end{array}\,\right)\left( \begin{array}{c}
 \eta_q\\
 \eta_s
 \end{array}\,\right).
 \ee

In this mixing scheme, we obtain the transition form factors
$F_{P\gamma}$ for $P\to\gamma^*\gamma ~(P=\pi^0, \eta, \eta')$ transitions
as follows
\bea\label{Eq5}
F_{\pi\gamma}(q^2) &=& \frac{(e^2_u - e^2_d)}{\sqrt{2}} I^{m_{u(d)}}_{\rm tot},
\nonumber\\
F_{\eta\gamma} (q^2) &=& \cos\phi\; \frac{(e^2_u + e^2_d)}{\sqrt{2}} I^{m_{u(d)}}_{\rm tot}
- \sin\phi\; e^2_s  I^{m_s}_{\rm tot},
\nonumber\\
F_{\eta'\gamma} (q^2) &=& \sin\phi\;\frac{(e^2_u + e^2_d)}{\sqrt{2}} I^{m_{u(d)}}_{\rm tot}
+ \cos\phi\; e^2_s  I^{m_s}_{\rm tot},
\eea
where we again should note that 
$I^{m_Q}_{\rm tot}=2[I^{m_Q}_{(a)}]^{\rm Cov}=2([I^{m_Q}_{(b)}]^{\rm LF}_{0<\al<1}+[I^{m_Q}_{(c)}]^{\rm LF}_{0<\al<1})
=2[I^{m_Q}_{(c)}]^{\rm LF}_{\alpha=1}=2[I^{m_Q}_{(b)}]^{\rm LF}_{\al=0}$ with the factor 2 needed
to include the contribution from the amplitude of the photon attached to the antiquark line.

\begin{figure}[h]
\centering
\includegraphics[width=0.8\columnwidth,clip=]{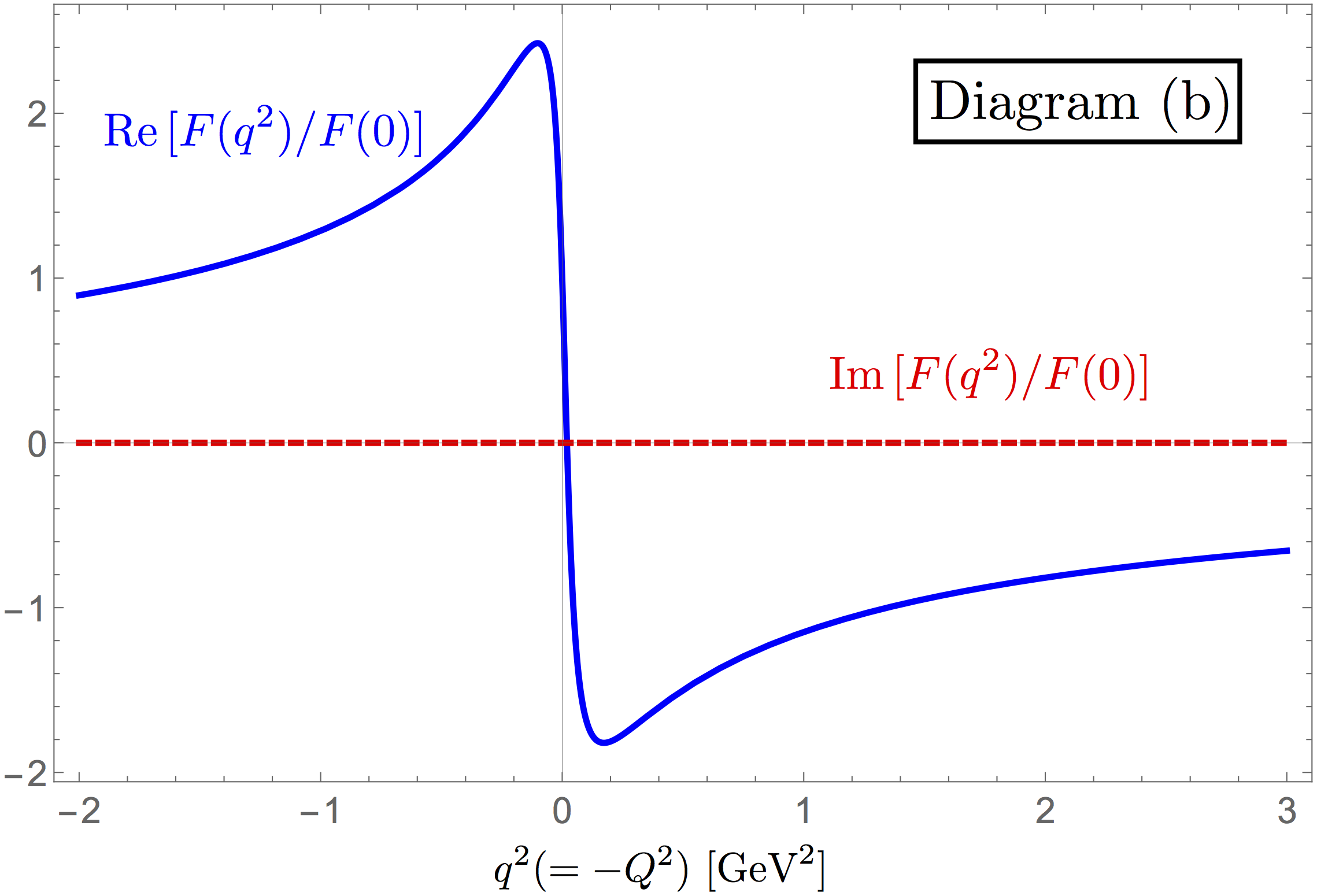}
\\
\centering {(a)}
\\
\includegraphics[width=0.8\columnwidth,clip=]{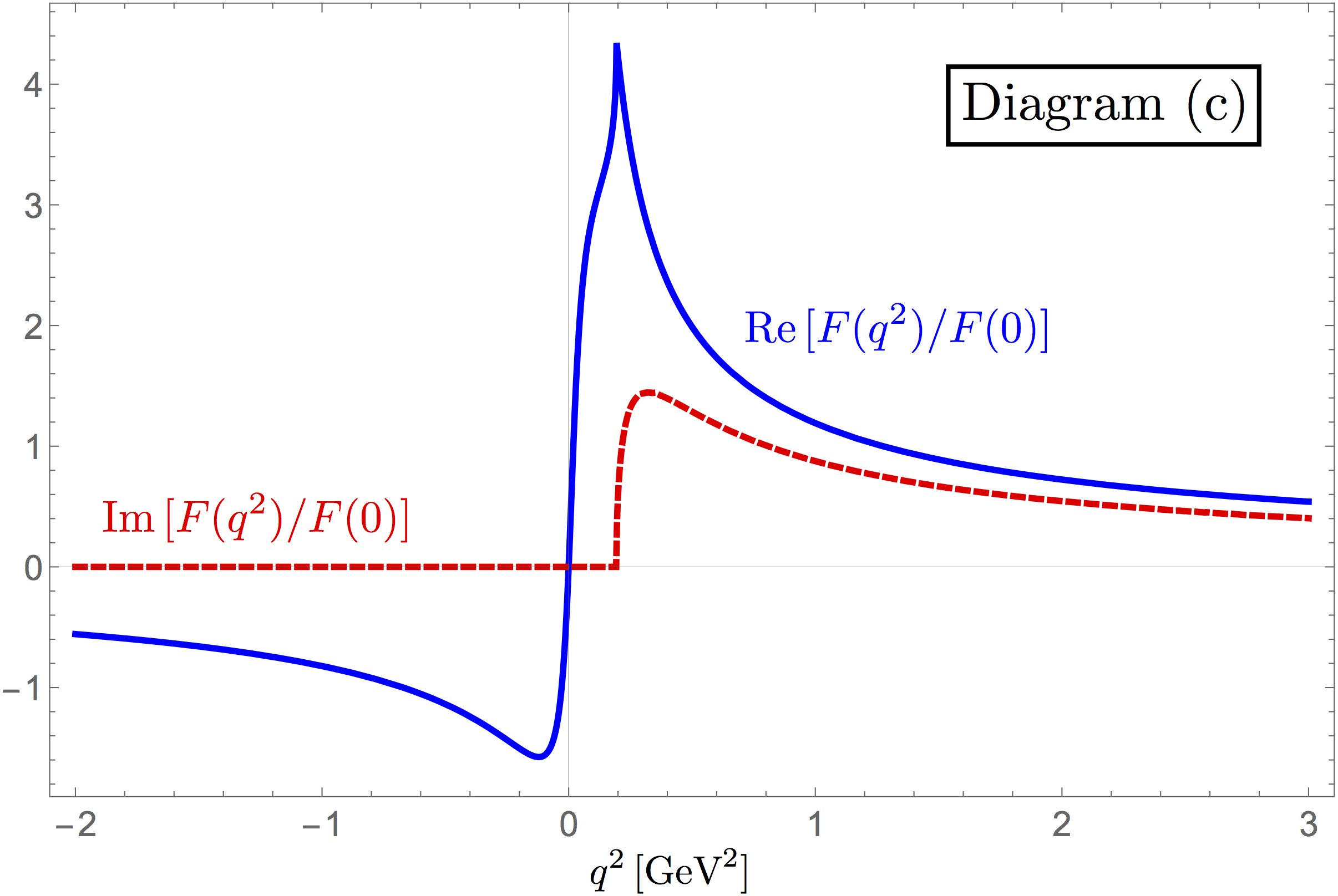}
\\
\centering {(b)}
\\
\includegraphics[width=0.8\columnwidth,clip=]{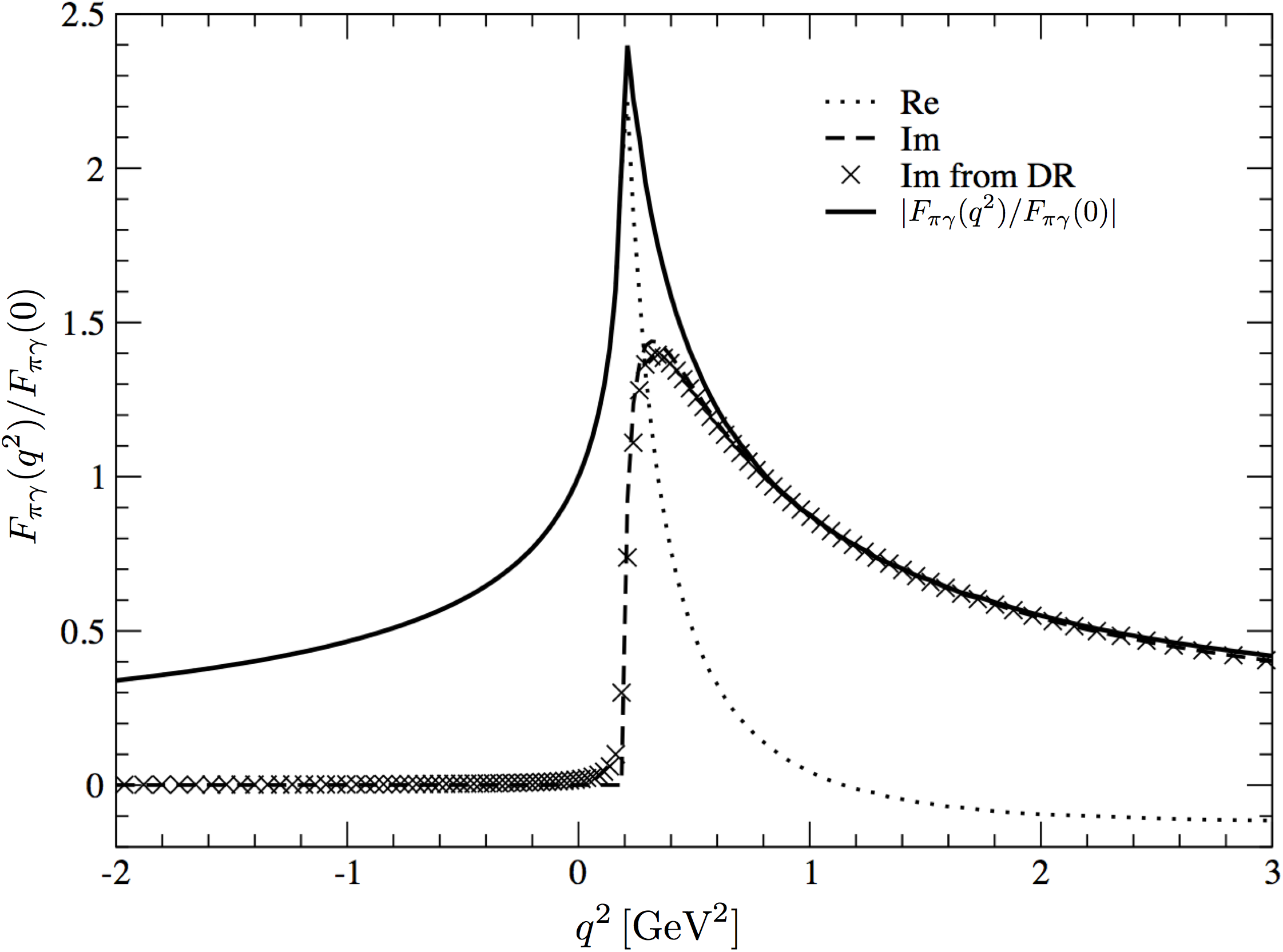}
\\
\centering{(c)}
%%%HMC
{\color{black}
\caption{\label{Nfig2}(Color online) The exactly solvable BS model calculation of the normalized $F_{\pi\gamma}(q^2)$ for both space- and time-like regions 
($-2\leq q^2\leq 3$ ) [GeV$^{2}$]: 
  (a) and~(b) represent the contributions from Figs.~\ref{fig1} (b) and~\ref{fig1} (c), respectively, for $0<\al<1$ case.
%$[I^{m_Q}_{(b)}]^{\rm LF}_{0<\al<1}$ in Eq.~(\ref{NEq7}) and  $[I^{m_Q}_{(c)}]^{\rm LF}_{0<\al<1}$ in Eq.~(\ref{NEq8}), respectively.  
 (c) shows the normalized $F_{\pi\gamma}(q^2)/F_{\pi\gamma}(0)$ compared with the dispersion relation. 
%(a), (b) and (c) 
%represent the results obtained from $[I^{m_Q}_{(b)}]^{\rm LF}_\al$, $[I^{m_Q}_{(c)}]^{\rm LF}_\al$ and 
%$[I^{m_Q}_{(a)}]^{\rm Cov}=[I^{m_Q}_{(b)}]^{\rm LF}_\al + [I^{m_Q}_{(c)}]^{\rm LF}_\al 
%=[I^{m_Q}_{(c)}]^{\rm LF}_{\alpha\to 1}=[I^{m_Q}_{(b)}]^{\rm LF}_{\al\to 0}$, respectively.
}}
\end{figure}

For the illustration of the numerical results from the exactly solvable BS model calculation, we show 
%%%HMC
{\color{black}
the normalized  pion TFF $F_{\pi\gamma}(q^2)/F_{\pi\gamma}(0)$  
}
for both space- and time-like regions
of momentum transfer  $-2\leq q^2\leq 3$ GeV$^2$ in Fig.~\ref{Nfig2}.
The used model parameters are $m_Q=0.22$ GeV and $M=0.14$ GeV.
%%%HMC
%%%CRJ
{\color{black} We note that the value of $g$ to yield 
the experimental data value for $F^{\rm Exp.}_{\pi\gamma}(0)=0.272$ GeV$^{-1}$ at $q^2=0$
is given by $g=3.22$.}
%%%CRJ
While $F_{\pi\gamma}(q^2)$ in spacelike momentum transfer region ($q^2 <0$) is real, it becomes complex in timelike region ($q^2>0$),
$F_{\pi\gamma}(q^2)= {\rm Re}\;F_{\pi\gamma}(q^2) + i {\rm Im}\;F_{\pi\gamma}(q^2)$.
Figs.~\ref{Nfig2} (a) and~\ref{Nfig2} (b) represent the results obtained from 
$[I^{m_Q}_{(b)}]^{\rm LF}_{0<\al<1}$ in Eq.~(\ref{NEq7}) and  $[I^{m_Q}_{(c)}]^{\rm LF}_{0<\al<1}$ in Eq.~(\ref{NEq8}), respectively.  As one can see, 
the imaginary part ${\rm Im}\;F_{\pi\gamma}(q^2)$ (red line) of the form factor  comes only from Fig.~\ref{fig1} (c) and starts to appear from the threshold $q^2=4m^2_Q$.
Fig.~\ref{Nfig2} (c) shows the normalized pion TFF,  $F_{\pi\gamma}(q^2)/F_{\pi\gamma}(0)$. 
%%%HMC
{\color{black}
The dotted, dashed and solid lines in Fig.~\ref{Nfig2} (c)
represent  ${\rm Re}\;[F_{\pi\gamma}(q^2)/F_{\pi\gamma}(0)]$, ${\rm Im}\;[F_{\pi\gamma}(q^2)/F_{\pi\gamma}(0)]$ 
and $|F_{\pi\gamma}(q^2)/F_{\pi\gamma}(0)|$, respectively.  
}
We confirmed numerically that 
$[I^{m_Q}_{(a)}]^{\rm Cov}=[I^{m_Q}_{(b)}]^{\rm LF}_{0<\al<1} + [I^{m_Q}_{(c)}]^{\rm LF}_{0<\al<1}
=[I^{m_Q}_{(c)}]^{\rm LF}_{\alpha = 1}=[I^{m_Q}_{(b)}]^{\rm LF}_{\al = 0}$ as mentioned earlier.
As a consistency check for our numerical calculations, we also compare our direct results 
of the form factor $F(q^2)= {\rm Re}\;F(q^2) + i {\rm Im}\;F(q^2)$ with those obtained from the dispersion
relations (DR) given by
\bea\label{ReF}
{\rm Re}\; F(q^2)&=&\frac{1}{\pi} P\int^\infty_{-\infty}\frac{ {\rm Im}\;F(q'^2)}{q'^2 -q^2} dq'^2,
\nonumber\\
{\rm Im}\; F(q^2)&=& -\frac{1}{\pi} P\int^\infty_{-\infty}\frac{ {\rm Re}\;F(q'^2)}{q'^2 -q^2} dq'^2,
\eea
where $P$ indicates the Cauchy principal value. 
In Fig.~\ref{Nfig2} (c), the data denoted by $(\times)$ represents  the DR result of  ${\rm Im}\;F(q^2)$ obtained from  Eq.~(\ref{ReF}) 
and shows an excellent agreement with  our direct result (dashed line) .
This assures the validity of our numerical calculation in the timelike region.

\section{Application of the Light-Front Quark Model}
\label{sec:III}

In our previous analysis of the twist-2 and twist-3 DAs of
pseudoscalar and vector mesons~\cite{TWV,TWPS,TWPS17} and the pion electromagnetic form factor~\cite{TWPS},
we have shown that standard LF (SLF) results of the LFQM is obtained by
the replacement of the LF vertex function $\chi$ in the BS model with the Gaussian wave function
$\phi_R$ as follows  [see, e.g., Eq. (35) in~\cite{TWPS}] 
\be\label{QM7}
 \sqrt{2N_c} \frac{ \chi(x,{\bf k}_\perp) } {1-x}
 \to \frac{\phi_R (x,{\bf k}_\perp) }
 {\sqrt{{\bf k}^2_\perp + m_Q^2}}, \; M \to M_0,
 \ee
where $M\to M_0$ implies that the physical mass $M$ included in the integrand of BS
amplitude (except $M$ in the vertex function $\chi$) has to be replaced with the invariant
mass $M_0$ since the SLF results of the LFQM
are obtained from the requirement of all constituents being on their respective mass-shell.
%%%CRJ
{\color{black} The mapping given by Eq.(\ref{QM7}) was originally found for the resolution of the LF zero-mode
issue in the vector meson decay constant and its self-consistent covariant description as
discussed extensively in~\cite{TWV}. As the mapping however involves only the radial wavefunction
and the meson mass, the same mapping holds for the pseudoscalar mesons as we have discussed
in~\cite{TWPS,TWPS17}. Likewise, the correspondence in Eq.~(\ref{QM7}) is valid again in this analysis of a $P \to \gamma^*\gamma$ transition.}
%%%CRJ

In the standard LFQM~\cite{CJ_PLB,CJ_99,CJ_DA,CJBc,Jaus90,Jaus91,CCP,Choi07} approach, the
wave function of a ground state pseudoscalar meson as a $q\bar{q}$ bound state is given by
\be\label{QM1}
\Psi_{\lam{\bar\lam}}(x,{\bf k}_{\perp})
={\phi_R(x,{\bf k}_{\perp})\cal R}_{\lam{\bar\lam}}(x,{\bf k}_{\perp}),
\ee
where $\phi_R$ is the radial wave function and the
spin-orbit wave function ${\cal R}_{\lam{\bar\lam}}$
with the helicity $\lam({\bar\lam})$ of a quark~(antiquark)
is obtained by the interaction-independent Melosh transformation~\cite{Melosh}
from the ordinary spin-orbit wave function assigned by the quantum numbers $J^{PC}$.

For the equal quark and antiquark mass $m_Q=m_{\bar Q}$,
the Gaussian wave function $\phi_R$ is given by
\be\label{QM2}
\phi_R(x,{\bf k}_{\perp})=
\frac{4\pi^{3/4}}{\beta^{3/2}}
\sqrt{\frac{M_0}{4 x (1-x)}} e^{m^2_Q/2\beta^2} e^{-M^2_0/8\beta^2},
\ee
where $\partial k_z/\partial x = M_0/4x(1-x)$ is the Jacobian of the variable transformation
$\{x,{\bf k}_\perp\}\to {\vec k}=({\bf k}_\perp, k_z)$
and $\beta$ is the variational parameter
fixed by our previous analysis of meson mass spectra~\cite{CJ_PLB,CJ_99,CJBc}.
The covariant form of the spin-orbit wave function ${\cal R}_{\lam{\bar\lam}}$
is given by
\be\label{QM4}
{\cal R}_{\lam{\bar\lam}}
=\frac{\bar{u}_{\lam}(p_Q)\gamma_5 v_{{\bar\lam}}( p_{\bar Q})}
{\sqrt{2}M_0},
\ee
and it satisfies
$\sum_{\lam{\bar\lam}}{\cal R}_{\lam{\bar\lam}}^{\dagger}{\cal R}_{\lam{\bar\lam}}=1$.
Thus, the normalization of our wave function is given by
\be\label{QM6}
1=
\int^1_0 dx \int\frac{d^2{\bf k}_\perp}{16\pi^3}
|\phi_R(x,{\bf k}_{\perp})|^2.
\ee
%That is, the probability $P_{q\bar q}$ of the valence Fock state of a meson
%is 1 in our LFQM.

Applying the correspondence given by Eq.~(\ref{QM7}) to  $[I^{m_Q}_{(c)}]^{\rm LF}_{\alpha = 1}$ in Eq.~(\ref{NEq10}) 
and $[I^{m_Q}_{(b)}]^{\rm LF}_{\al = 0}$ in Eq.~(\ref{Eq8}),
we obtain the corresponding SLF results 
$[I^{m_Q}_{\rm tot}]^{\rm SLF}_{\alpha = 1}$
and
$[I^{m_Q}_{\rm tot}]^{\rm SLF}_{\al = 0}$
in our LFQM as follows:
\be\label{QM8a}
[I^{m_Q}_{\rm tot}]^{\rm SLF}_{\alpha = 1} =\frac{\sqrt{2 N_c}}{4\pi^3}\int^{1}_0
 \frac{dx}{(1-x)} \int d^2{\bf k}_\perp
 \frac{m_Q}{M^2_0 - q^2}
 \frac{\phi_R(x,{\bf k}_{\perp})}{\sqrt{{\bf k}^2_\perp + m^2_Q}},
 \ee
 and 
\be\label{QM8b}
[I^{m_Q}_{\rm tot}]^{\rm SLF}_{\al = 0} = \frac{\sqrt{2 N_c}}{4\pi^3}\int^{1}_0
 \frac{dx}{x} \int d^2{\bf k}_\perp
  \frac{m_Q}{M^{\prime 2}_0}
 \frac{\phi_R(x,{\bf k}_{\perp})}{\sqrt{{\bf k}^2_\perp + m^2_Q}},
\ee
respectively. We confirm that our result is frame-independent, i.e.,
$[I^{m_Q}_{\rm tot}]^{\rm SLF}_{0<\alpha<1}=[I^{m_Q}_{\rm tot}]^{\rm SLF}_{\alpha = 1}=[I^{m_Q}_{\rm tot}]^{\rm SLF}_{\al = 0}$.
While the TFFs for $P\to\gamma^*\gamma$ can be obtained by substituting either
$[I^{m_Q}_{\rm tot}]^{\rm SLF}_{\al = 1}$  or $[I^{m_Q}_{\rm tot}]^{\rm SLF}_{\al = 0}$
into Eq.~(\ref{Eq5}), we shall use $[I^{m_Q}_{\rm tot}]^{\rm SLF}_{\al = 1}$ for the analysis of the timelike region 
due to the simple and clean pole structure given by $(M^2_0 - q^2)^{-1}$ in~Eq.~(\ref{QM8a}) compared to the
pole appearing through $[M^{\prime 2}_0]^{-1}$ in~Eq.~(\ref{QM8b}). It is important to notice that 
the internal transverse momentum ${\bf k}_\perp$ doesn't mix with the external virtual photon momentum
$q=(P^+, q^2/P^+, 0)$ in $\alpha = 1$ case as shown in Eq.(\ref{QM8a}) so that the direct timelike TFF calculation can be done most effectively.
For sufficiently high spacelike momentum transfer $Q^2(=-q^2={\bf q}^2_\perp)$
region, both Eqs.~(\ref{QM8a}) and~(\ref{QM8b})  can be approximated
in the leading order of $1/Q^2$ as follows
\be\label{QM9}
[I^{m_Q}_{\rm tot}]^{\rm SLF} \simeq 2 f_P \int^{1}_0 \frac{dx}{(1-x)Q^2}
 \int d^2{\bf k}_\perp \psi_P (x,{\bf k}),
 \ee
where $f_P$ is the pseudoscalar meson decay constant and $\psi_P (x, {\bf k}_\perp)$
is the transverse momentum dependent DA (TMDA)~\cite{Rad14}
that is a 3-dimensional generalization of the twist-2 pseudoscalar meson DA $\phi_{2;P}(x)$:
\bea\label{2DA}
\phi_{2;P} (x) &=& \frac {\sqrt{2N_c}} {f_P 8\pi^3} \int d^2{\bf k}_\perp
\frac{\phi_{R}(x, {\bf k}_\perp)}{\sqrt{ {\bf k}^2_\perp + m_Q^2} }{m_Q}
\nonumber\\
&=& \int^\infty_0  d^2{\bf k}_{\perp} \psi_P(x, {\bf k}_\perp).
\eea
%The first and second terms in the curly bracket of Eq.~(\ref{QM9}) correspond to the contributions of the leading order
%and next-to-leading order in $1/Q$  to the TFFs, respectively.
%As we show in our numerical calculation,
%our TFFs are applicable to the timelike region by analytically continuing the spacelike $Q^2\to -Q^2$ in Eq.~(\ref{QM8}) or~(\ref{QM9}).

%Furthermore, Eq.~(\ref{QM9}) is exactly the same as previous SLF result
From Eqs.~(\ref{Eq5}), (\ref{QM9}) and~(\ref{2DA}), one can verify that our LFQM result for 
$F_{\pi\gamma}(Q^2)$ at sufficiently high $Q^2$ can be approximated as
\be\label{QM10}
F_{\pi\gamma}(Q^2)\simeq \frac{\sqrt{2} f_\pi}{3}
\int^{1}_0 \frac{dx}{(1-x)Q^2}\phi_{2;\pi} (x).
\ee
The asymptotic PQCD DA, $\phi_{2;\pi} (x) = 6 x(1-x)$, leads to the well-known Brodsky-Lepage limit~\cite{BL80}: $Q^2 F_{\pi\gamma}(Q^2\to\infty)=\sqrt{2}f_{\pi}\simeq 0.185$ GeV.
%As we shall show in the numerical section, our LFQM result for $F^{\rm %LFQM}_{\pi\gamma}(Q^2)$ satisfies the asymptotic limit set by perturbative QCD:
%$Q^2F_{\pi\gamma}(Q^2)=\sqrt{2}f_\pi\simeq 0.185$ GeV.

%\section{Decay Widths for $P\to\gamma\gamma$ Transition}
Applying our LFQM to calculate the decay widths for
$P\to\gamma\gamma~(P=\pi^0,\eta, \eta')$
transition, the decay width for $P\to\gamma\gamma$ is given by
\be\label{DW1}
\Gamma_{P\to\gamma\gamma} =\frac{\pi}{4}\alpha^2M^3|F_{P\gamma}(0)|^2,
\ee
where $\alpha$ is the fine structure constant.
{\color{black}
The form factor $F_{P\gamma}(0)$ at $Q^2=0$ 
}
may also be
expressed in terms of the decay constants obtained 
%by the axial-vector anomaly plus
%PCAC (partial conservation of axial-vector current)
%In the chiral limit, the form factor zero may be obtained 
from the Adler-Bell-Jackiw (ABJ) anomaly (or the chiral anomaly)~\cite{ABJ1,ABJ2} as follows
\bea\label{DW2}
F^{\rm ABJ}_{\pi\gamma}(0)&=& \frac{1}{2\sqrt{2}\pi^2 f_\pi},
\nonumber\\
F^{\rm ABJ}_{\eta\gamma}(0)&=& \frac{1}{2\sqrt{6}\pi^2}
\biggl[\frac{1}{f_8}\cos\theta-\frac{2\sqrt{2}}{f_0}\sin\theta\biggr],
\nonumber\\
F^{\rm ABJ}_{\eta'\gamma}(0)&=& \frac{1}{2\sqrt{6}\pi^2}
\biggl[\frac{1}{f_8}\sin\theta+\frac{2\sqrt{2}}{f_0}\cos\theta\biggr],
\eea
where $\theta$ is the mixing angle in the flavor SU(3) octet-singlet basis and
is related with the mixing angle $\phi$ in the quark-flavor basis via
$\theta=\phi - {\rm arctan}\sqrt{2} \simeq \phi - 54.7^\circ$.
While the quadratic (linear) Gell-Mann-Okubo mass formula prefers
$\theta\simeq -10^\circ, \phi\simeq 44.7^\circ$
($\theta\simeq -23^\circ, \phi\simeq 31.7^\circ$),
the KLOE Collaboration~\cite{KLOE} extracted the pseudoscalar mixing
angle $\phi$
by measuring the ratio ${\rm BR}(\phi\to\eta'\gamma)/{\rm BR}(\phi\to\eta\gamma)$.
The measured values are $\phi=(39.7\pm 0.7)^{\circ}$ and
$(41.5\pm 0.3_{\rm stat}\pm 0.7_{\rm syst}\pm 0.6_{\rm th})^{\circ}$
with and without the gluonium content for $\eta'$, respectively.
The mixing angle has also been analyzed on lattice by RBC-UKQCD Collaboration~\cite{RBC},
where  $\theta=-14.1(2.8)^\circ$ was obtained.
However, since the mixing angle for $\eta-\eta'$ is still a controversial issue, we
use more conservatively $\phi=37^\circ\pm 5^\circ$ to check the sensitivity of our LFQM.

\section{Numerical Results}
\label{sec:IV}

\begin{table}[t]
\caption{The constituent quark masses $m_Q(Q=u(d), s)$ (in GeV) and the Gaussian parameters
$\beta_{Q{\bar Q}}$ (in GeV) for the linear confining potentials
obtained from the variational principle in our LFQM~\cite{CJ_PLB,CJ_99,CJ_DA}.}
\label{t1}
\begin{tabular}{cccc} \hline\hline
 $m_{u(d)}$ & $m_s$ & $\beta_{Q{\bar Q}}$ & $\beta_{s{\bar s}}$ \\
\hline
 0.22~ & 0.45~ & 0.3659~ & 0.4128~  \\
%HO & 0.25 &  0.48 & 0.3194 & 0.3419  \\
\hline\hline
\end{tabular}
\end{table}
\begin{table}
\caption{
{\color{black}
Form factor $F_{P\gamma}(0)$
 (in $\rm{GeV}^{-1}$)} for $(P=\pi^0,\eta,\eta')\to\gamma\gamma$.}
\label{t2}
\begin{tabular}{lcccc} \hline\hline
Model & $F_{P\gamma}^{\rm Th.}(0)$ & $F^{\rm ABJ}_{P\gamma}(0)$ & $F^{\rm Exp.}_{P\gamma}(0)$  \\
\hline
$\pi\to\gamma\gamma$ &  0.242 & 0.276 & $0.272(3)$   \\
\hline
%\hspace{1.05cm} ($\phi=32^\circ$) & $0.xxx$ &  $0.261$ & $0.274(5)$  \\
$\eta\to\gamma\gamma$ (for $\phi=37^{-5^\circ}_{+5}$) & $0.286^{+0.024}_{-0.027}$ &  $0.232^{+0.029}_{-0.030}$ & $0.274(5)$  \\
%\hspace{1.05cm} (for $\phi=42^\circ$) & $0.259$ &  $0.202$ & $0.274(5)$  \\
\hline
$\eta'\to\gamma\gamma$ (for $\phi=37^{-5^\circ}_{+5}$) & $0.290^{-0.026}_{+0.024}$ &  $0.332^{-0.021}_{+0.025}$ & 0.344(6) \\
%\hspace{1.12cm} (for $\phi=42^\circ$) & 0.314 &  0.357 & 0.344(6)  \\
\hline\hline
\end{tabular}
\end{table}
\begin{figure*}
\vspace{0.5cm}
\begin{center}
\includegraphics[height=7cm, width=8cm]{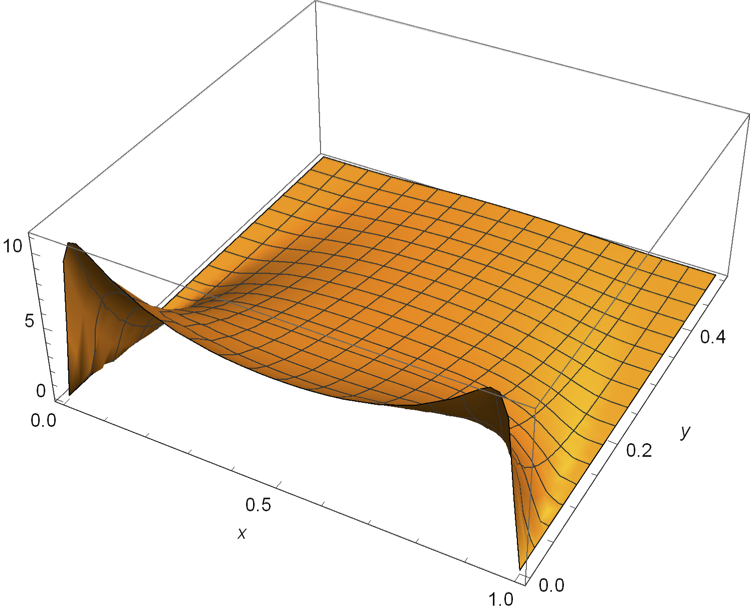}
\hspace{0.5cm}
%\\
\includegraphics[height=7cm, width=8cm]{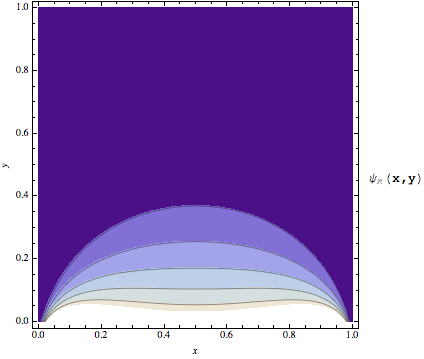}
\caption{\label{Nfig3} (Color online) Transverse momentum dependent distribution amplitude (TMDA)
$\psi_\pi (x, {\bf k}_\perp)$ (left panel) for the pion in the form of 
{\color{black}
$\psi_\pi(x,y)$ [see Eq.~(\ref{TMDA2})] 
}
and the corresponding two-dimensional contour plot (right panel). }
\end{center}
\end{figure*}
\begin{figure*}
\vspace{0.5cm}
\begin{center}
\includegraphics[height=7cm, width=8cm]{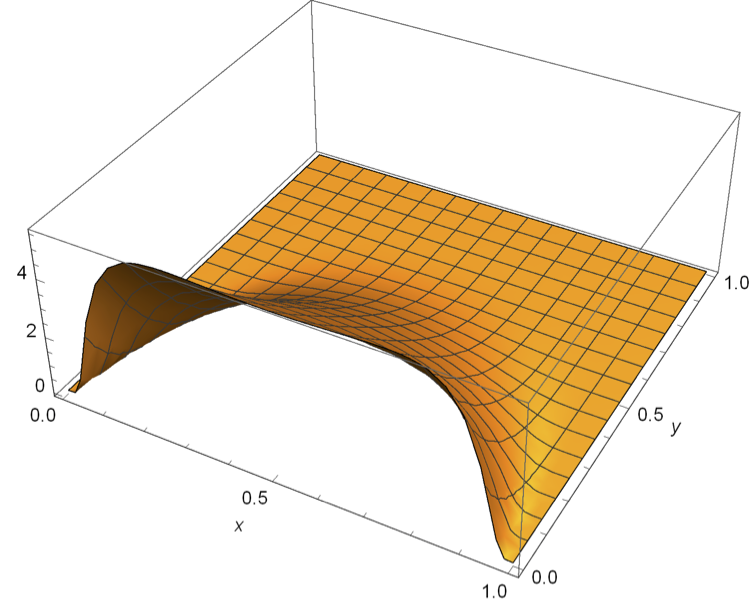}
\hspace{0.5cm}
%\\
\includegraphics[height=7cm, width=8cm]{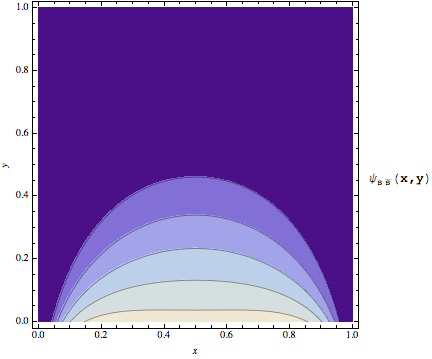}
\caption{\label{Nfig4} (Color online) Transverse momentum dependent distribution amplitude (TMDA)
$\psi_{s\bar{s}}(x, {\bf k}_\perp)$ (left panel) for the $s\bar{s}$ sector 
in the form of
{\color{black}
 $\psi_{s\bar{s}}(x,y)$ [see Eq.~(\ref{TMDA2})] 
 }
and the corresponding two-dimensional contour plot (right panel). }
\end{center}
\end{figure*}

In our numerical calculations within the standard LFQM, we use the model parameters
(i.e. constituent quark masses $m_Q$ and the gaussian parameters $\beta_{Q{\bar Q}}$) for the
linear confining potentials given in Table~\ref{t1}, which were obtained from the calculation of meson mass spectra using 
the variational principle in our LFQM~\cite{CJ_PLB,CJ_99,CJ_DA}.
For the model parameters given in Table~\ref{t1}, our LFQM predictions of 
the decay constants for the pion, octet ($\eta_8$) and singlet ($\eta_0$) mesons are
$f_{\pi}=130$ MeV, $f_8/f_\pi = 1.32$, and $f_0/f_\pi=1.16$, respectively.
Our results of the octet and singlet meson decay constants are quite comparable
with other theoretical predictions such as 
 $f_8/f_\pi=1.26$ and $f_0/f_\pi=1.17$~\cite{FKS},
 $f_8/f_\pi=1.28$ and $f_0/f_\pi=1.25$~\cite{Leut98},
 and $f_8/f_\pi=1.25$ and $f_0/f_\pi=1.04\pm 0.04$~\cite{chi}.

For the numerical computations of the TFFs given by Eq.~(\ref{Eq5}) using our LFQM,
we use the result $[I^{m_Q}_{\rm tot}]^{\rm SLF}_{\alpha = 1}$ in Eq.~(\ref{QM8a})
since it is much more convenient to handle the singularities in timelike momentum transfer region than any other reference
frame.
In Table~\ref{t2}, we summarize our LFQM results of
{\color{black}
 form factor $F_{P\gamma}(0)$
for $(P=\pi^0,\eta,\eta')\to\gamma\gamma$ 
}
obtained from the direct calculation~[$F_{P\gamma}^{\rm Th.}(0)$]
(see Eqs.~(\ref{Eq5})~and~(\ref{QM8a}) ) and from the ABJ formulae~[$F^{\rm ABJ}_{P\gamma}(0)$] (see Eq.~(\ref{DW2}))
compared with the experimental data~\cite{PDG,KLOE2}. For the
$(\eta,\eta')\to\gamma\gamma$ processes, we use the mixing angles
$\phi=37^{-5^\circ}_{+5}$ in the quark-flavor basis.
The experimental values of $F^{\rm Exp.}_{\eta\gamma}(0)=0.274(5)$ GeV$^{-1}$
and $F^{\rm Exp.}_{\eta'\gamma}(0)=0.344(6)$ GeV$^{-1}$ were extracted from
the measured decay widths $\Gamma^{\rm Exp.}(\eta\to\gamma\gamma)=0.516(18)$ keV
(obtained after combining the PDG average~\cite{PDG} together with the recent KLOE-2
result~\cite{KLOE2})
and $\Gamma^{\rm Exp.}(\eta'\to\gamma\gamma)=4.35(14)$ keV, respectively.
For the $\pi^0\to\gamma\gamma$ case, while our result $F^{\rm ABJ}_{\pi\gamma}(0)=0.276$ {\color{black} GeV$^{-1}$}
obtained from ABJ anomaly
is in good agreement with the data, 
%$F^{\rm Exp}_{\pi\gamma}(0)=0.272(3)$, 
the direct result $F_{\pi\gamma}^{\rm Th.}(0)=0.242$ {\color{black} GeV$^{-1}$}
accounts for about 90$\%$ of the data.
For the $(\eta,\eta')\to\gamma\gamma$ case, while our results $F^{\rm ABJ}_{P\gamma}(0)$
prefer $\phi\simeq 32^\circ$
to fit the data, the direct results $F_{P\gamma}^{\rm Th.}(0)$ prefer
$\phi\simeq 40^\circ$ for the best fits of both $\eta$ and $\eta'$ TFFs.

From the point of view of QCD, the twist-2 DA $\phi_{2;P}(x)$ of a hadron depends on the scale $\mu$ which separates 
nonperturbative and perturbative regimes. In our LFQM, we can associate $\mu$ with the transverse integration cutoff 
via $|{\bf k}_\perp|\leq \mu$, which is the usual way how the normalization scale is defined for the LF wave function~(see, e.g. Ref.~\cite{BL80}). 
In order to estimate this cutoff value, we made a three-dimensional plot for TMDA $\psi_P(x, {\bf k}_\perp)$ in Eq.~(\ref{2DA})
in the form of $\psi_P(x,y)$ by changing the variable ${\bf k}^2_\perp = y / (1-y)$
{\color{black}
so that
%$y$ ranges from 0 to 1. 
\be\label{TMDA2}
\phi_{2;P}(x) = \int^\infty_0 d^2 {\bf k}_\perp\;\psi_P(x, {\bf k}_\perp) 
=\int^1_0\;dy\;\psi_P (x,y),
\ee
where $\psi_P (x,y)=\pi \psi_P(x, |{\bf k}_\perp|=\sqrt{y/(1-y)})/(1-y)^2$.
}

Fig.~\ref{Nfig3} shows the three-dimensional plot ({\color{black} left panel}) of $\psi_\pi(x,y)$ for the pion 
and the corresponding two-dimensional contour plot (right panel). In fact, we obtain the twist-2 pion DA $\phi_{2;\pi}(x)$ by performing 
the transverse integration up to infinity (or equivalently $y$ up to 1) without loss of accuracy due to the presence of Gaussian 
damping factor. However, we find that the integration up to $y \simeq 0.5$~(or equivalently
$\mu\simeq |{\bf k}_\perp|\simeq 1$ GeV) of $\psi_\pi(x, y)$ makes up $99\%$ of the full result for $\phi_{2;\pi}(x)$. This implies that 
our cutoff scale corresponds to
$\mu\simeq |{\bf k}_\perp|\simeq 1$ GeV for the calculation of the twist-2 $\phi_{2;\pi}(x)$.
The twist-2 and twist-3 DAs for $\pi$ can be found in our previous works~\cite{CJ_DA, TWPS,TWPS17}.
In Fig.~\ref{Nfig4}, we show the three-dimensional plot (left panel) of 
{\color{black} $\psi_{s\bar{s}}(x,y)$}
for the $s\bar{s}$ sector 
and the corresponding two-dimensional contour plot (right panel). 
In the case of $s\bar{s}$ sector, the cutoff scale corresponds to $\mu\simeq 1.13$ GeV.

\begin{figure}
%\vspace{0.5cm}
\begin{center}
\includegraphics[height=7cm, width=8cm]{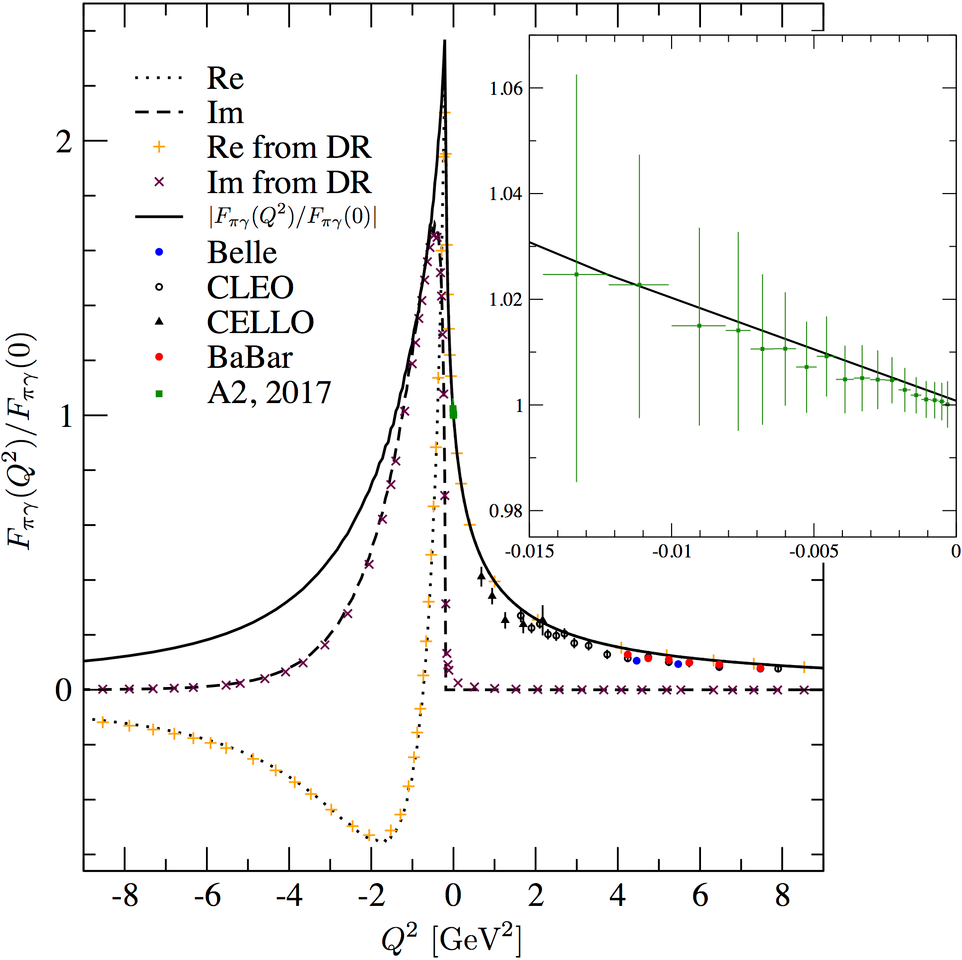}
%\vspace{1.2cm}
\\
\centering{(a)}
\\
\includegraphics[height=7cm, width=8cm]{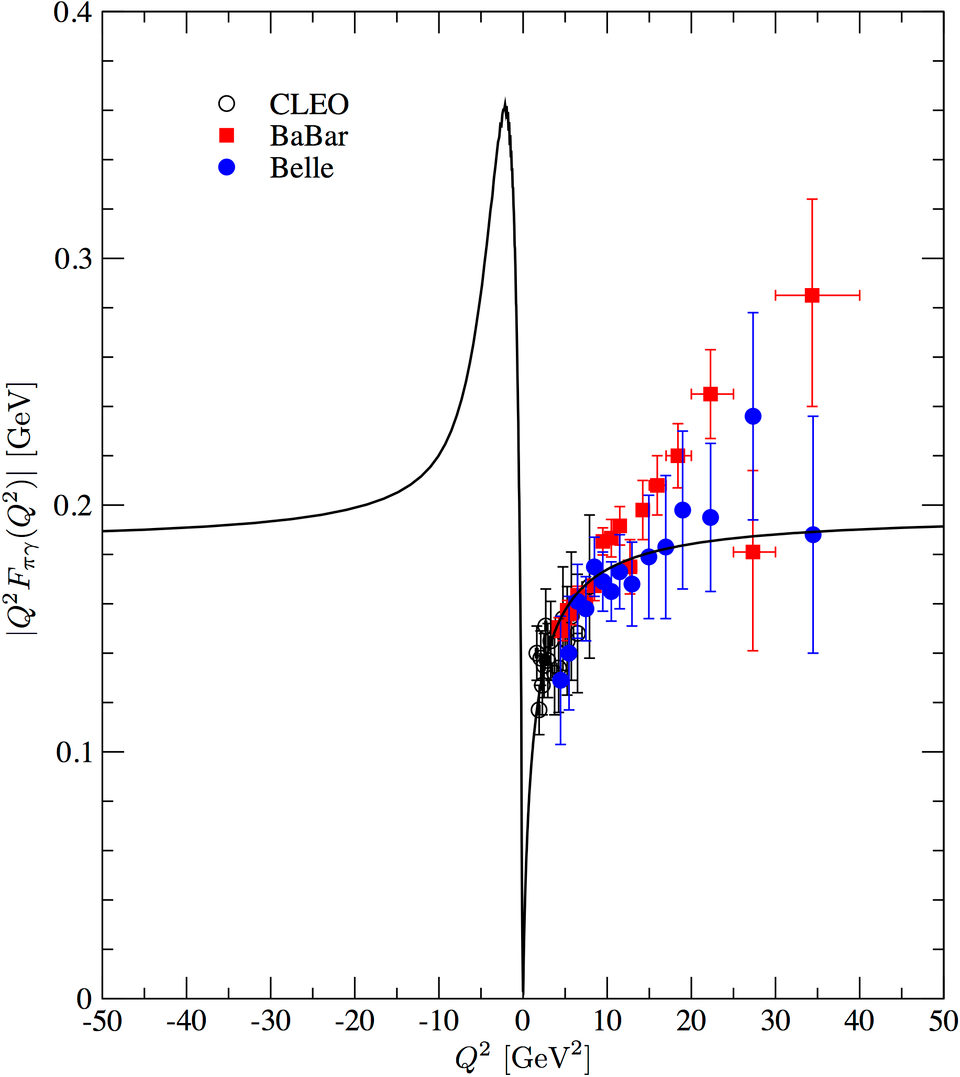}
\\
\centering{(b)}
\caption{\label{NFig5} (Color online) (a) The the normalized $\pi\to\gamma^*\gamma$ transition
form factor
$F_{\pi\gamma}(Q^2)/F_{\pi\gamma}(0)$, and (b) {\color{black} $|Q^2 F_{\pi\gamma}(Q^2)|$ }
for both timelike ($q^2=-Q^2>0$) and spacelike ($q^2=-Q^2<0$) momentum transfer regions. 
The data are taken from~\cite{CELLO91,CLEO98,A2pi,Babar09,Belle12} and~\cite{A2pi}. }
\end{center}
\end{figure}
In Fig.~\ref{NFig5}, we show the normalized $\pi^0\to\gamma^*\gamma$ TFF
$F_{\pi\gamma}(Q^2)/F_{\pi\gamma}(0)$ (see Fig.~\ref{NFig5} (a)) and $|Q^2 F_{\pi\gamma}(Q^2)|$ (see  Fig.~\ref{NFig5} (b))
for both timelike ($q^2=-Q^2>0$) and spacelike ($q^2=-Q^2<0$) momentum transfer region 
using Eqs.~(\ref{Eq5}) and~(\ref{QM8a}) and compare them with the available experimental data
for the spacelike region~\cite{CELLO91,CLEO98,Babar09,Belle12} and for the small timelike region ($0<q^2< m^2_{\pi}$ GeV$^2$)
obtained from the pion Dalitz decay~\cite{A2pi}. 
{\color{black}
The dotted, dashed and solid lines in Fig.~\ref{NFig5} (a) represent our LFQM predictions
of  ${\rm Re}\;[F_{\pi\gamma}(q^2)/F_{\pi\gamma}(0)]$, ${\rm Im}\;[F_{\pi\gamma}(q^2)/F_{\pi\gamma}(0)]$ 
and $|F_{\pi\gamma}(q^2)/F_{\pi\gamma}(0)|$, respectively.  
}
%The solid, dotted and dashed lines in  Fig.~\ref{NFig5} (a) represent our LFQM predictions
%of the total result, the real part and the imaginary part of the normalized form factor $F_{\pi\gamma}(Q^2)/F_{\pi\gamma}(0)$, respectively.
We note that the spacelike  region can
be easily obtained by analytically continuing the momentum transfer $q^2\to -q^2$ in the integrand of Eq.~(\ref{QM8a}).
As one can see from Fig.~\ref{NFig5},
our result for low- and intermediate- spacelike $Q^2$ region show a good agreement with the data.
%In our previous calculation~\cite{CJCS} of the pion electromagnetic
%form factor in both spacelike- and timelike region, we  also explicitly showed that the timelike form factor obtained by the analytic
%continuation of the spacelike form factor exhibits complex structure and satisfies the dispersion relation.

As a consistency check of our LFQM calculations for the timelike region, 
we also include the real (imaginary) part of the form factor  obtained 
from the DR (denoted by $+(\times)$ data points) given by Eq. ~(\ref{ReF}). 
As one can see, our direct results for  the real and  imaginary parts are in perfect agreement with the results obtained from the DR. 
While the exactly solvable BS model calculation shows the dominant contribution of ${\rm Im}\;F_{\pi\gamma}$ for most of the
timelike region, the LFQM result of ${\rm Im}\;F_{\pi\gamma}$ with the more realistic Gaussian radial wave function
shows dominant contribution only near the resonance region and the  timelike region above $q^2>1$ GeV$^2$ is dominated by the real part contribution. 
That is, the relative contribution between the real and imaginary parts depends on the shape of the hadron bound state wave function.
We also note that Kroll~\cite{Kroll11} 
made a rough estimate for the expected size of the timelike form factor using the modified perturbative approach (MPA)~\cite{MPA1,MPA2},
i.e.,
the timelike form factor is dominantly real for $q^2$ larger than 5 GeV$^2$, its imaginary part contributes less than about $10\%$
to the absolute value. Kroll's discussion about the relative strength between the real and imaginary parts is qualitatively consistent with our LFQM prediction and the reason for this may be attributed to the usage of similar type of the Gaussian wave function.

As one can see from Fig.~\ref{NFig5} (a),  our result for the small timelike region is in good agreement with 
the very recent measurement of the $\pi^0\to e^+ e^-\gamma$ Dalitz decay from the A2 Collaboration~\cite{A2pi}.  
The slope parameter can be defined from  
the vector meson dominance (VMD) model in which  the normalized TFF is typically  parametrized as~\cite{PDG}
\be\label{VMD}
F_P(m_{ll}) = \frac{1}{1- \frac{m_{ll}^2}{\Lambda^2_P}} \simeq 1 + a_P \frac{m^2_{ll}}{m^2_{P}} \, ,
\ee
where $m_{ll}= \sqrt{q^2} $ is the dilepton invariant mass and $a_P=(m_P/\Lambda_P)^2$  reflects the form-factor slope
at $q^2=0$. Our result for the slope parameter $a_\pi$ for the $\pi^0$ TFF is obtained as 
\be\label{api_LFQM}
a_\pi = 0.0355,
\ee
%Our result 
which shows a good agreement with 
%the two recent experimental data,
the current world average $a_\pi = 0.032 \pm 0.004$~\cite{PDG} obtained from
timelike measurements~\cite{api1,api2,api3} and the extrapolation of spacelike
data~\cite{CELLO91} using a VMD model, as well as the two recent experimental data extracted from the $\pi^0\to e^+e^-\gamma$ Dalitz decay, 
$a_\pi=0.030\pm 0.010$ from A2 Collaboration~\cite{A2pi}
and $a_\pi=0.0368\pm 0.0057$ from NA62 Collaboration~\cite{NA60-17}.
%as well as the value of Review of Particle
%Physics~\cite{PDG},  $a_\pi=0.032\pm 0.004$ determined by a measurement of the spacelike $\pi^0$ TFF in the 
%$e^+e^-\to e^+e^-\pi^0$ process by CELLO~\cite{CELLO91}.
Our result should also be compared with other theoretical predictions: 
$a_\pi=0.0288(42)$ from a Lattice QCD with two flavors of quarks~\cite{Lattice16};
$a_\pi=0.0324(12)_{\rm stat}(19)_{\rm syst}$ from the method of Pad\'e approximants~\cite{Mas12}; 
$a_{\pi}=0.032(1)$ from  a Regge analysis~\cite{AB10}; 
$a_\pi=0.036$ from the ChPT~\cite{BBC};
$a_\pi=0.029(5)$ from a study of the Dalitz decay of $\pi^0$~\cite{Kampf}; 
$a_\pi\approx 0.031$~\cite{GR08} and $a_\pi\approx 0.035$~\cite{DGH08} from a hard-wall holographic model of QCD; 
and $a_\pi =0.024(5)$~\cite{Col12} from a soft-wall holographic model of QCD.
For the analysis of  timelike form factor near resonance region in Fig.~\ref{NFig5} (a),
the maximum value of  $F_{\pi\gamma}(q^2)$  occurs  at  $q^2\simeq 4m^2_Q$ due to the virtual photon wave function term
$1/(M^2_0 -q^2)$ in Eq.~(\ref{QM8a}). Since the peak position of the timelike TFF in our LFQM depends on the value of the constituent 
quark mass, the $\rho$-pole type resonance may be obtained by simply taking $m_{u(d)}=M_\rho/2$. 
%Such $\rho$-pole type resonance may also be achieved by finding more realistic form of the photon wave function than the above one,
%but this is beyond the scope of the present work.

Fig.~\ref{NFig5} (b) shows $|Q^2 F_{\pi\gamma}(Q^2)|$ for the extensive range ($-50 \leq Q^2 \leq 50$ GeV$^2$) of both time- and space-like
momentum transfer regions compared with the spacelike experimental data~\cite{CLEO98,Babar09,Belle12}.
We note that our LFQM result for $|Q^2 F_{\pi\gamma}(Q^2)|$  for the spacelike region $10 \leq Q^2\leq 45$ GeV$^2$
is in good agreement with the data from Belle~\cite{Belle12} showing the asymptotic behavior
but disagree with the BaBar data~\cite{Babar09} showing the rapid growth for this $Q^2$ regime.
In our LFQM calculation for the perturbative region, we find slightly different values  for the timelike and spacelike TFFs,
e.g. we find the absolute values of 
{\color{black}
$|Q^2F_{\pi\gamma}(Q^2)| \simeq 0.194$ GeV in the spacelike region
and 
$|q^2F_{\pi\gamma}(q^2)| \simeq 0.186$ GeV in the timelike region
at $|Q^2|=112$ GeV$^2$, respectively.
}
%, and the corresponding ratio of the spacelike to timelike TFFs is about 1.04.
%This result is in agreement with the conclusion in \cite{BRS00} that perturbative corrections 
%cannot generate a significant difference between the spacelike and timelike TFFs.
%We obtain the asymptotic result as
%$\lim_{Q\to\infty}Q^2F_{\pi\gamma}(Q^2)\simeq 0.195$ GeV, which is
%about $5\%$ larger than 
%Our results for $|Q^2|=112$ GeV$^2$ value are quite comparable with the asymptotic limit set by perturbative QCD:
%$Q^2F_{\pi\gamma}(Q^2)=\sqrt{2}f_\pi\simeq 0.185$ GeV~\cite{BL80}.
%This difference may be a guide to estimate the valence $q\bar{q}$ state contribution
%(e.g. in this work, we set the probability of finding $q\bar{q}$ state in a
%meson as $P_{q\bar{q}}=1$) and the higher-Fock state contribution.
Although there may be some contributions from the higher-twist
and higher Fock-state as discussed in~\cite{BCT},
however, we infer  from the results shown in Fig.~\ref{NFig5} that
the higher Fock-state contribution may not be large, especially, for high $Q^2$ regime.

\begin{figure}
%\vspace{0.5cm}
\begin{center}
\includegraphics[height=7cm, width=8cm]{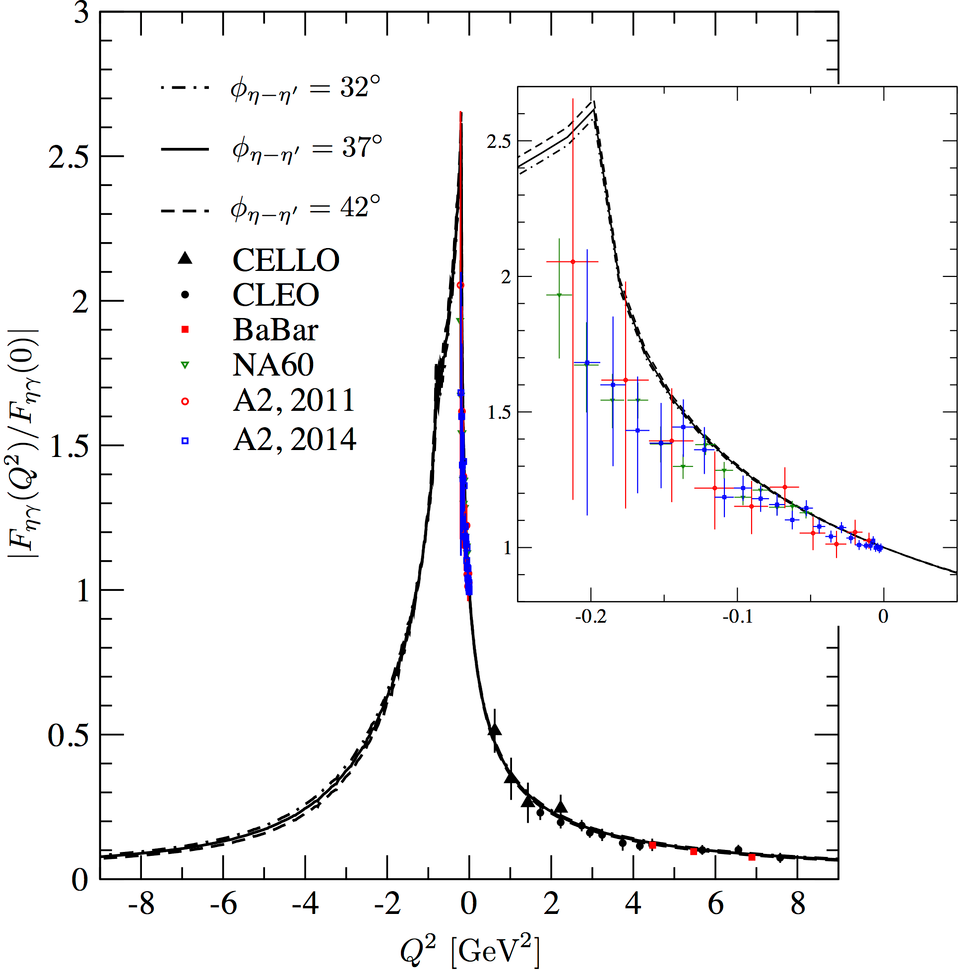}
%\vspace{1.2cm}
\\
\centering{(a)}
\\
\includegraphics[height=7cm, width=8cm]{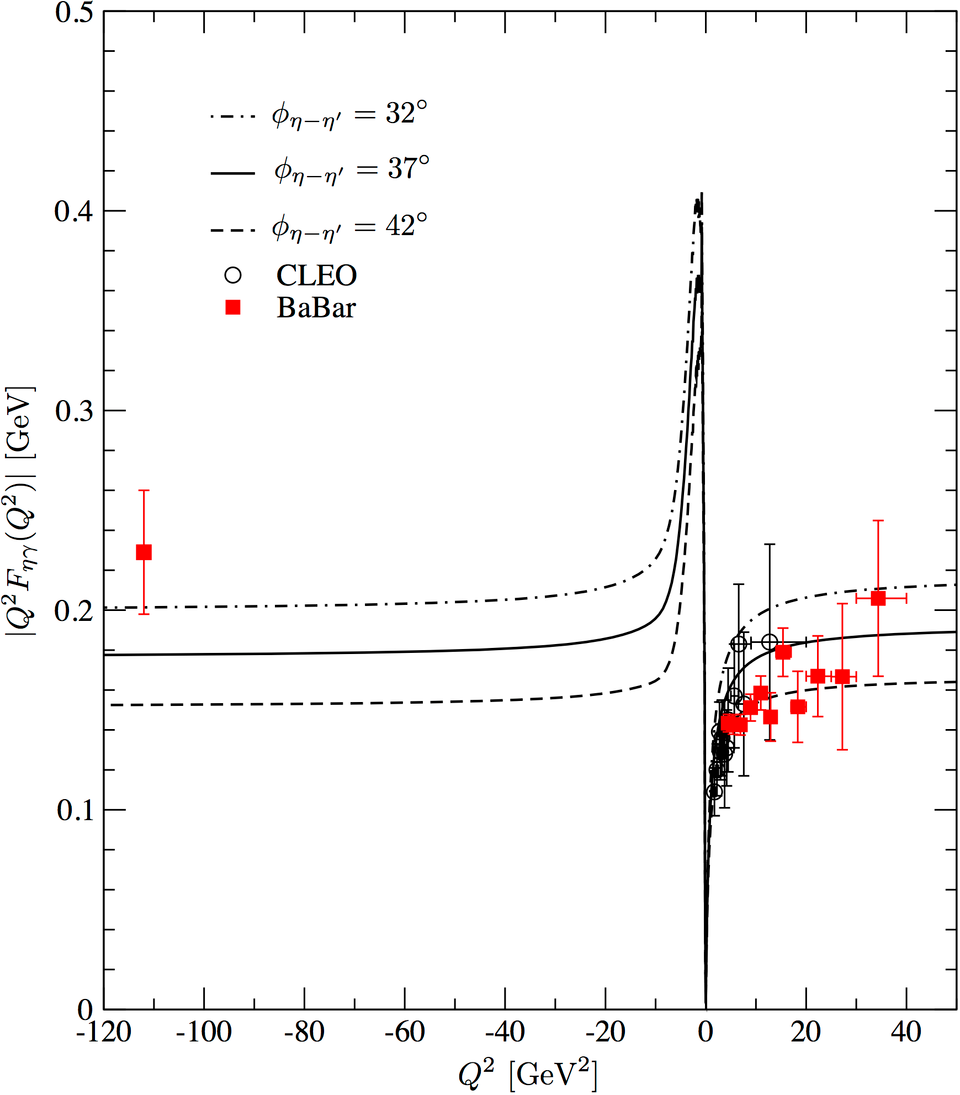}
\\
\centering{(b)}
\caption{\label{NFig6} (a) The normalized $\eta\to\gamma^*\gamma$ transition
form factor
{\color{black}
$|F_{\eta\gamma}(Q^2)/F_{\eta\gamma}(0)|$, and (b)  $|Q^2 F_{\eta\gamma}(Q^2)|$
 } 
for both timelike ($q^2=-Q^2>0$) and spacelike ($q^2=-Q^2<0$) momentum transfer region. 
The dot-dashed, solid, and dashed lines are results obtained from the mixing angles with
$\phi_{\eta-\eta'}=32^\circ$, $37^\circ$ and $42^\circ$, respectively, and
the data are taken from~\cite{CELLO91,CLEO98,Babar11,A22014,A22011,NA60}.  }
\end{center}
\end{figure}

\begin{figure}
%\vspace{0.5cm}
\begin{center}
\includegraphics[height=7cm, width=8cm]{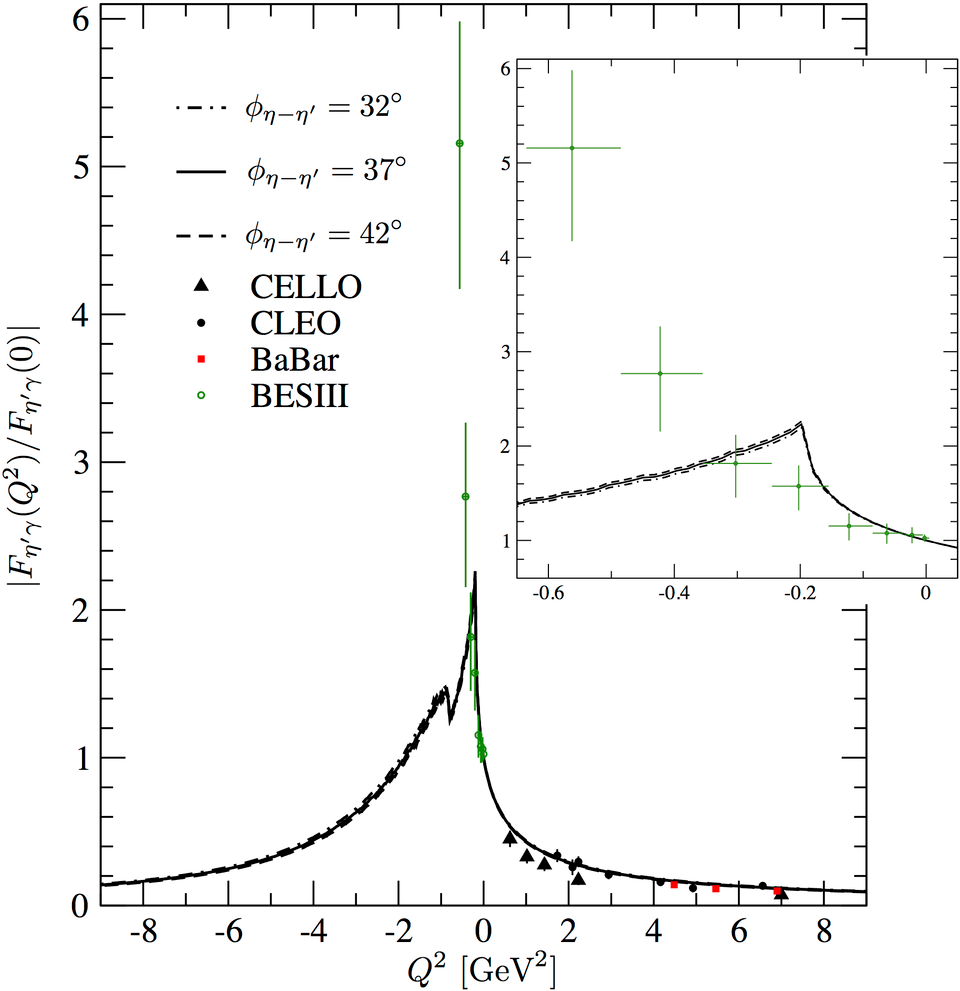}
%\vspace{0.1cm}
\\
\centering{(a)}
\\
\includegraphics[height=7cm, width=8cm]{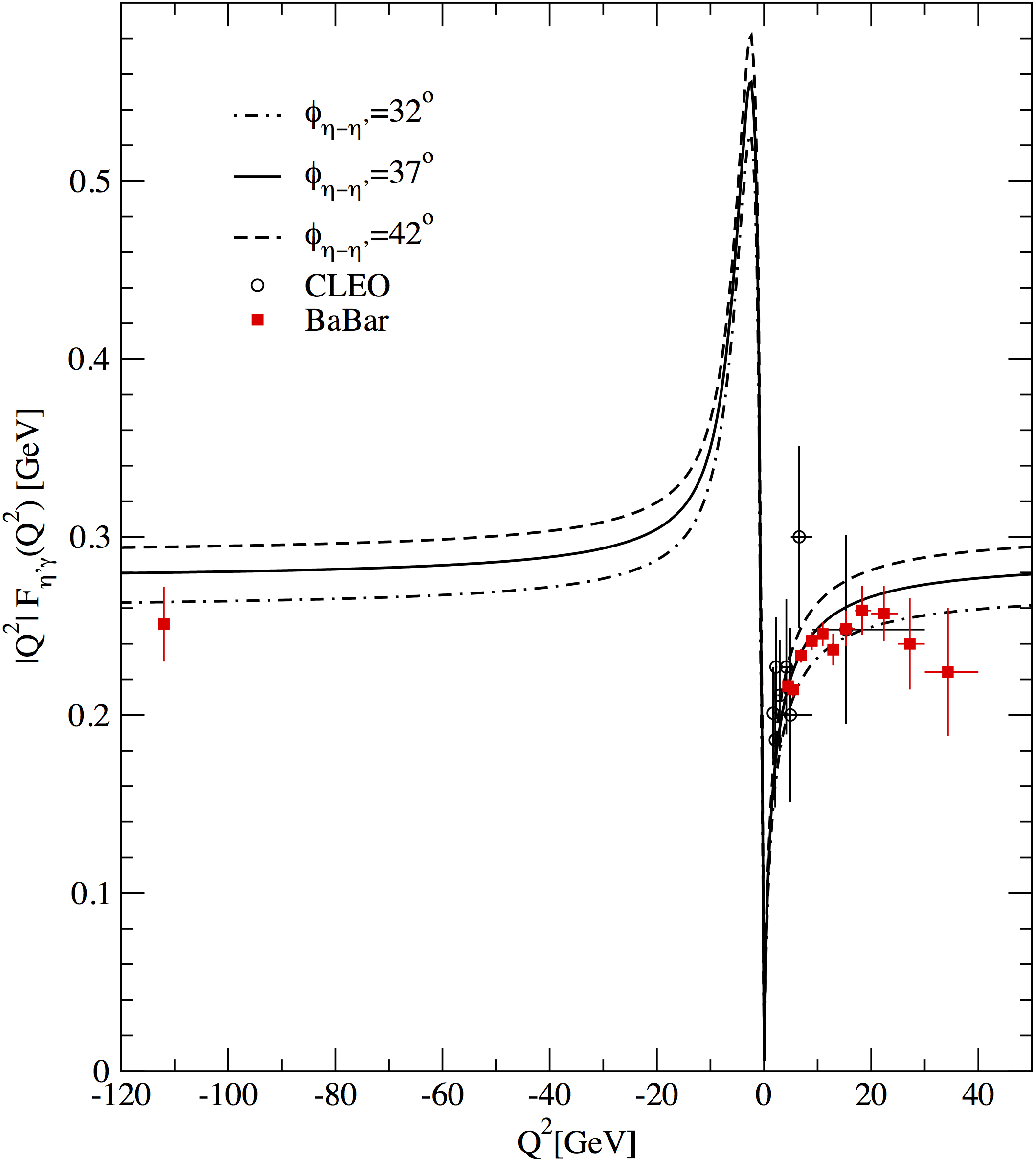}
\\
\centering{(b)}
\caption{\label{NFig7}  (a) The normalized $\eta'\to\gamma^*\gamma$ transition
form factor
{\color{black}
$|F_{\eta'\gamma}(Q^2)/F_{\eta'\gamma}(0)|$, and (b)  $|Q^2 F_{\eta'\gamma}(Q^2)|$
 }
for both timelike ($q^2=-Q^2>0$) and spacelike ($q^2=-Q^2<0$) momentum transfer region. 
The same line codes are used as in Fig.~\ref{NFig6} and
the data are taken from~\cite{BES15,CELLO91,CLEO98,Babar11}. }
\end{center}
\end{figure}

In Fig.~\ref{NFig6}, we show the normalized $\eta\to\gamma^*\gamma$ TFF
{\color{black} 
$| F_{\eta\gamma}(Q^2)/F_{\eta\gamma}(0) |$ (see Fig.~\ref{NFig6} (a)) and $|Q^2 F_{\eta\gamma}(Q^2)|$  
}
(see Fig.~\ref{NFig6} (b))
for both time- and space-like momentum transfer region. 
The corresponding figures for $\eta'$ TFFs are shown in Fig.~\ref{NFig7}.
Since the patterns for the real and imaginary parts of the $\eta$ and $\eta'$ TFFs
are similar to those of $\pi^0$ TFF, we only show
the total results  for the $\eta$ and $\eta'$ TFFs but varying the mixing angles.
Since they are rather sensitive to the $\eta-\eta'$ mixing angles, we display the results
with the variation of the mixing angles as a sensitivity check.
The dot-dashed, solid, and dashed lines 
%in Figs.~\ref{NFig6} and~\ref{NFig7} are total results 
{\color{black} for $|F_{\eta\gamma}(q^2)/F_{\eta\gamma}(0)|$  in Fig.~\ref{NFig6}
and $|F_{\eta'\gamma}(q^2)/F_{\eta'\gamma}(0)|$ in Fig.~\ref{NFig7}
are 
}
obtained from the mixing angles with
$\phi_{\eta-\eta'}=32^\circ$, $37^\circ$ and $42^\circ$, respectively. 
The experimental data for spacelike region
are taken from~\cite{CELLO91,CLEO98,Babar11}.
The small timelike data in Figs.~\ref{NFig6} (a)  and~\ref{NFig7} (a) are taken  from the 
measurements of the $\eta$  ($0\leq q^2\leq M^2_\eta$ GeV$^2$) and $\eta'$  ($0\leq q^2\leq M^2_{\eta'}$ GeV$^2$) Dalitz decays;
$\eta\to\ell{\bar\ell}\gamma(\ell=e,\mu)$~\cite{NA60,A22014,A22011} and $\eta'\to e^+e^-\gamma$~\cite{BES15}.

For the small and medium momentum transfer  in both timelike and spacelike regions (i.e. $-8\leq Q^2\leq 8$ GeV$^2$) as shown
in Figs.~\ref{NFig6} (a) and ~\ref{NFig7} (a), both normalized TFFs $F_{\eta\gamma}$ and $F_{\eta'\gamma}$ 
are not sensitive to the the variation of the the mixing angles $\phi_{\eta-\eta'}=(37 \pm 5)^\circ$ and
show good agreement with the available data in spacelike region~\cite{CELLO91,CLEO98,Babar11}.
For the comparison with the timelike data from the $(\eta,\eta')$ Dalitz decays, we obtain 
the slope parameters $\Lambda^{-2}_{\eta(\eta')}=a_{\eta(\eta')}/m^2_{\eta(\eta')}$ defined by Eq.~(\ref{VMD}) as follows 
\bea\label{etaslope}
\Lambda^{-2}_\eta &=& 2.112^{-0.031}_{+0.038}\; {\rm GeV}^{-2}\;\;
{\rm for}\;\phi=37^{-5^\circ}_{+5},
\nonumber\\
\Lambda^{-2}_{\eta'} &=& 1.732^{-0.035}_{+0.031}\; {\rm GeV}^{-2}\;\;
{\rm for}\;\phi=37^{-5^\circ}_{+5}.
\eea
which correspond to $\Lambda_\eta = 688^{+5}_{-6}$ MeV and $\Lambda_{\eta'}= 760^{+8}_{-7}$ MeV
for $\phi=37^{-5^\circ}_{+5}$, respectively.
Our results for the slope parameters for $\eta$ and $\eta'$ TFFs are consistent with the available 
experimental data within the error bars:  
$\Lambda^{-2}_\eta=(1.95\pm 0.22)$ GeV$^{-2}$~\cite{NA60} and $(1.95\pm 0.25)$ GeV$^{-2}$~\cite{A22014} for $\eta$ TFF, and 
$\Lambda^{-2}_{\eta'}=(1.60\pm 0.16)$ GeV$^{-2}$~\cite{CELLO91},
$(1.6\pm 0.25)$ GeV$^{-2}$~\cite{BES15}, and 
$(1.7\pm 0.4)$ GeV$^{-2}$~\cite{Lepton-G} for $\eta'$ TFF, respectively.
We also should note that
the ratio of $\Lambda_{\eta'}$ to $\Lambda_{\eta}$  is insensitive to the mixing angle, i.e. $\Lambda_{\eta'}/\Lambda_\eta\simeq 1.11$
for $32^\circ\leq \phi \leq 42^\circ$.

For the resonance properties of $F_{\eta\gamma}$ and $F_{\eta'\gamma}$ within our LFQM as shown in Figs.~\ref{NFig6} (a) and ~\ref{NFig7} (a),
the primary and secondary peaks of  both $F_{\eta\gamma}(q^2)$ and $F_{\eta'\gamma}(q^2)$
occurs at $q^2\simeq 4m_Q^2(Q=u,d)$ and $q^2\simeq 4m_s^2$, respectively, regardless of their mixing angles. 
That is, the $\eta-\eta'$ mixing effect is not significant for the small timelike region corresponding to the $\eta$- and $\eta'$- Dalitz decays.
Particularly, the secondary peak for $F_{\eta'\gamma}$ is more pronounced than that for 
%%%HMC
$F_{\eta\gamma}$. 
This may
be ascribed to the fact that $F_{\eta'\gamma}$ receives contribution more from $s\bar{s}$ component than $Q\bar{Q}(Q=u,d)$ components. 
%%%%
For this kinematic regions of the $\eta$- and $\eta'$ Dalitz decays, 
while our LFQM result for the $F_{\eta\gamma}$ is quite comparable with the data~\cite{A22014,A22011,NA60},
our result for $F_{\eta'\gamma}$ shows large deviation from the recent BESIII data~\cite{BES15} except near $q^2=0$ region.
This large deviation for $F_{\eta'\gamma}$ near $q^2=M^2_\rho$ may be expected from the property of our LFQM, in which the
primary peak appears at $q^2=4m^2_{u(d)}$ rather than $q^2=M^2_\rho$. We expect from our LFQM analysis that the experimental data 
for both timelike $F_{\eta\gamma}$ and $F_{\eta'\gamma}$ would show peaks near $q^2=M^2_\rho$ 
and $q^2=M^2_\phi$ corresponding to our primary and secondary peaks at $q^2=4m^2_{u(d)}$ and $q^2=m^2_s$, respectively.

While the mixing angle effects on $F_{\eta\gamma}$ and $F_{\eta'\gamma}$ do not appear too significant for small and medium 
momentum transfer region (i.e. $|Q^2|<8$ GeV$^2$)  as shown in Figs.~\ref{NFig6} (a) and ~\ref{NFig7} (a),
its effects become substantial for large momentum transfer region (i.e. $|Q^2|>10$ GeV$^2$) as shown in Figs.~\ref{NFig6} (b) and ~\ref{NFig7} (b). 
%Our prediction $F_{\eta\gamma}(Q^2)$
%using both mixing angles $\phi=37^\circ$ and~$42^\circ$ show good agreement with the available data not only for the
%low $Q^2$ region but also for the high $Q^2$ region. 
As in the case of $\pi^0\to\gamma^*\gamma$
transition, our predictions for  both 
{\color{black}
$|Q^2 F_{\eta\gamma}(Q^2)|$ and
 $|Q^2 F_{\eta'\gamma}(Q^2)|$ 
}
 show asymptotic behavior for
$|Q^2|\geq 40$ GeV$^2$ region. 
The single timelike data at $q^2=s=112$ GeV$^2$  in Fig.~\ref{NFig6} (b) and the one in Fig.~\ref{NFig7} (b)
are taken from the measurement of $e^+ e^-\to\gamma^*\to \eta(\eta')\gamma$ process 
at the center of mass of  $\sqrt{s}=10.58$ GeV by the BaBar Collaboration~\cite{Babar11}:
\bea\label{Q2etap}
s|F_{\eta\gamma}| &=& (0.229 \pm 0.031) \; {\rm GeV},
\nonumber\\
s|F_{\eta'\gamma}| &=& (0.251 \pm 0.021) \; {\rm GeV}.
\eea

In our LFQM calculation for the perturbative region, we find slightly different values  for the timelike and spacelike $\eta$ and $\eta'$ TFFs,
e.g. while the absolute spacelike values at $Q^2=112$ GeV$^2$ are 
{\color{black}
$|Q^2F_{\eta(\eta')\gamma}(Q^2)|\simeq 0.191^{+0.024}_{-0.025}( 0.284^{-0.017}_{+0.016}) $ GeV,
the timelike value at $q^2=112$ GeV$^2$ are
$|q^2F_{\eta(\eta')\gamma}(q^2)|\simeq 0.178^{+0.024}_{-0.025} ( 0.280^{-0.017}_{+0.016})$ GeV 
}
for  $\phi=37^{-5^\circ}_{+5}$, respectively.
But the corresponding ratios of the spacelike to timelike $\eta$ and  $\eta'$ TFFs at $|Q^2|=112$ GeV$^2$ are about 1.07 and 1.02, respectively,
regardless of the mixing angles.
Our results at the timelike $q^2=112$ GeV$^2$ are also consistent with the perturbative QCD predictions~\cite{Kroll11}, where
{\color{black}
$|q^2F_{\eta\gamma}(q^2)| \simeq 0.17$ GeV 
and $|q^2F_{\eta'\gamma}(q^2)| \simeq 0.28$ GeV were obtained.
}
As stated in~\cite{eta3}, while the BaBar result for $q^2F_{\eta\gamma}(q^2)$ at $q^2=112$ GeV$^2$ is about 2$\sigma$ 
larger than the asymptotic prediction, the corresponding result for $q^2F_{\eta'\gamma}(q^2)$ from the BaBar Collaboration
is in agreement with the asymptotic expectation.
Thus, it is hard to estimate the correct $\eta-\eta'$ mixing angle with these two experimental data points at $q^2=112$ GeV$^2$ in the present time.
More experimental data in perturbative region may be necessary to draw any definite conclusion on
the mixing angle.

\section{Summary and Discussion}
\label{sec:V}
In this work, we investigated the $(\pi^0,\eta,\eta')\to\gamma^*\gamma$ transitions for the entire kinematic regions analyzing both spacelike and timelike TFFs in the standard LF (SLF) approach within the phenomenologically accessible realistic LFQM~\cite{CJ_PLB,CJ_99,CJ_DA,CJBc}.  Performing the LF calculation in the covariant BS model as the first illustration, 
we used three different reference frames, i.e. (1) $q^+\neq 0$ frame with $P^- =q^-$ and $\al=q^+/P^+=q^2/M^2$, (2) $q^+\neq 0$ frame with $P^+=q^+$ (i.e.,
$\alpha = 1$), and (3) $q^+=0$ frame with $q^2=-{\bf q}^2_\perp=-Q^2$ (i.e, $\al = 0$), and found that all three different reference LF frames give
exactly the same results to the one obtained from the manifestly covariant calculation as they must be.
Especially, the calculation of $F_{P\gamma}$ using the $q^+\neq 0$ frame with $\al = 1$ is found to be most effective for the analysis of the timelike region due to the absence of mixing between the internal transverse momentum and the external virtual photon momentum that leads to the very simple pole structure $1/(q^2 - M^2_0)$ in the form factor.  
We also confirmed the absence of the LF zero mode in pseudoscalar meson TFFs. As a consequence,
the $q^+=0$ frame (i.e. $\al = 0$) calculation exhibits that the meson TFF using the plus component of the current is immune to the zero mode. 
Thus, in the $q^+=0$ frame (i.e., the well-known Drell-Yan-West frame), the complete total amplitude is provided by just the valence contribution
depicted in Fig.~\ref{fig1} (b).
As a consistency check for our numerical calculations, we also compared our direct results 
of the form factor $F(q^2)= {\rm Re}\;F(q^2) + i {\rm Im}\;F(q^2)$ with those obtained from the dispersion
relations (DR) and found the excellent agreement between the two results. 
This assured the reliability of our numerical calculation in the timelike region.

% As the SLF approach within the LFQM by itself is not amenable to determine the zero-mode contribution, 
%we utilized the covariant BS model to check the existence (or absence) of the zero mode.

We then mapped this exactly solvable manifestly covariant BS model to the standard LFQM following the same correspondence relation Eq.~(\ref{QM7}) between 
the two models that we found in our previous analysis of two-point and three-point functions for the pseudoscalar and vector 
mesons~\cite{TWV,TWPS}. This allowed us to apply the more phenomenologically accessible 
Gaussian wave function provided by the LFQM analysis of meson mass spectra~\cite{CJ_PLB,CJ_99,CJ_DA,CJBc}. 
In the analysis of the meson-photon TFFs using our LFQM, we took the $q^+\neq 0$ frame with $\al = 1$ which is the most convenient frame to analyze the timelike region compare to any other reference frames. For the $(\eta,\eta')\to\gamma^*\gamma$ transitions, we used the $\eta-\eta'$
mixing angle $\phi$ in the quark-flavor basis varying the $\phi$ values in the range of $\phi=37^{+5^\circ}_{-5}$ to check the sensitivity of our LFQM.
For the numerical analyses of the $P\to\gamma^*\gamma \;(P=\pi^0,\eta,\eta')$ TFFs using our LFQM, 
we investigated both the low-energy and high-energy regimes.

For the low-energy regime, our results for the TFFs and their slope parameters are in good agreement with the available data
from the Dalitz decays of $(\pi^0,\eta,\eta')$ mesons.
Especially, in the low momentum transfer region, the $\eta$ and $\eta'$ TFFs are rather insensitive to the mixing angles.
For the analysis of  timelike form factor near resonance region,
the maximum value of  $F_{\pi\gamma}$  occurs  at  $q^2(=-Q^2)\simeq 4m^2_Q$ due to the virtual photon wave function term
$1/(M^2_0 -q^2)$ in Eq.~(\ref{QM8a}). The $\rho$-pole type resonance may be achieved by finding more realistic form of 
the photon wave function, which is open for the future work.
For the resonance properties of $F_{\eta\gamma}$ and $F_{\eta'\gamma}$,
the primary and secondary peaks of  both $F_{\eta\gamma}$ and $F_{\eta'\gamma}$
occurs at $q^2\simeq 4m_Q^2(Q=u,d)$ and $q^2\simeq 4m_s^2$, respectively, regardless of their mixing angles.
We also anticipate from our LFQM analysis that the experimental data 
for both timelike $F_{\eta\gamma}$ and $F_{\eta'\gamma}$ would show peaks near $q^2=M^2_\rho$ (primary)
and $q^2=M^2_\phi$ (secondary) corresponding to our primary and secondary peaks at $q^2=4m^2_{u(d)}$ and $m^2_s$, respectively.

For the high-energy regime, our result of 
{\color{black}
$|Q^2 F_{P\gamma}(Q^2)|$ 
}
does not show any steep rising  behavior for high $|Q^2|$ region 
as measured from the BaBar Collaboration~\cite{Babar09} but shows scaling behavior for high $|Q^2|$ consistent with the 
perturbative QCD prediction. This is ascribed to the fact that our twist-2 DA~\cite{CJ_DA,TWPS} is highly
suppressed at the end points ($x=0,1$) unlike the flat DA~\cite{Ra09,MP09} showing the
enhancement at the end points. 
Especially, in our LFQM calculation for the perturbative region, we find slightly different values  for the timelike and spacelike TFFs,
e.g.  the ratios of the spacelike to timelike TFFs at $|Q^2|=112$ GeV$^2$ are about 1.04 for $\pi^0$ TFF and 1.07 (1.02) for $\eta(\eta')$,
regardless of the $\eta-\eta'$ mixing angles.
While the BaBar result~\cite{Babar11} for 
{\color{black}
$|q^2F_{\eta\gamma}(q^2)|$ 
}
at $q^2=112$ GeV$^2$ is about 2$\sigma$ 
larger than the asymptotic prediction, the corresponding result for 
{\color{black}
$|q^2F_{\eta'\gamma}(q^2)|$}
from the BaBar Collaboration
is in agreement with the asymptotic expectation.
Thus, it is hard to predict the correct $\eta-\eta'$ mixing angle with these two experimental data points at $q^2=112$ GeV$^2$ at present time.
More experimental data in perturbative region may be necessary to draw any definite conclusion on the mixing angle.

%%%CRJ
{\color{black} While the pseudoscalar meson vertex $\Gamma_P=\gamma_5$ is taken in this work,
the generalization of the vertex including the axial vector coupling~\cite{Pham} may be considered for further study. The work along this direction is underway.}  

\acknowledgments
H.-M. Choi was supported by the National Research Foundation of Korea (NRF)
(Grant No. NRF-2017R1D1A1B03033129). H.-Y. Ryu was supported by the NRF grant funded by
the Korea government(MSIP) (No. 2015R1A2A2A01004238).
C.-R. Ji was supported in part by the US Department of Energy
(Grant No. DE-FG02-03ER41260).

\section*{APPENDIX: $\eta-\eta'$ mixing}
In this appendix, we provide the comparison of the
$\eta-\eta'$ mixing angle between the octet-singlet basis and quark-flavor basis.
The octet-singlet mixing angle $\theta$ of $\eta$ and $\eta'$
is known to be in the range of $-10^{\rm o}$ to $-23^{\rm o}$~\cite{PDG}.
The physical $\eta$ and $\eta'$ are the mixtures of the flavor $SU(3)$ octet
$\eta_8$ and singlet $\eta_0$ states:
 \be\label{eet}
 \left( \begin{array}{cc}
 \eta\\
 \eta'
 \end{array}\,\right)
 =\left( \begin{array}{cc}
 \cos\theta\;\; -\sin\theta\\
 \sin\theta\;\;\;\;\;\cos\theta
 \end{array}\,\right)\left( \begin{array}{c}
 \eta_8\\
 \eta_0
 \end{array}\,\right),
 \ee
 where
$\eta_8=(u\bar{u}+d\bar{d}-2s\bar{s})/\sqrt{6}$ and
$\eta_0=(u\bar{u}+d\bar{d} + s\bar{s})/\sqrt{3}$. Analogously,
in terms of the quark-flavor(QF) basis $\eta_q=(u\bar{u}+d\bar{d})/\sqrt{2}$ and
 $\eta_s=s\bar{s}$, one obtains~\cite{FKS}
  \be\label{eea}
 \left( \begin{array}{cc}
 \eta\\
 \eta'
 \end{array}\,\right)
 =\left( \begin{array}{cc}
 \cos\phi\;\; -\sin\phi\\
 \sin\phi\;\;\;\;\;\cos\phi
 \end{array}\,\right)\left( \begin{array}{c}
 \eta_q\\
 \eta_s
 \end{array}\,\right).
 \ee
 The two schemes are equivalent to each
 other by $\phi=\theta+ \arctan\sqrt{2}$ when ${\rm SU}_f(3)$ symmetry is perfect.
Although it was
frequently assumed that the decay constants follow the same
pattern of state mixing, the mixing properties of
the decay constants will generally be different from those of the meson state since the
decay constants only probe
the short-distance properties of the valence Fock states while the
state mixing refers to the mixing of the overall wave
function~\cite{FKS}.

Defining
$\la P(p)|J^{q(s)}_{\mu5}|0\ra = -if^{q(s)}_P p^\mu$ ($P=\eta,\eta'$)
in the QF basis, the
four parameters $f^{q}_P$ and $f^{s}_P$ can be expressed in terms of two
mixing angles ($\phi_q$ and $\phi_s$)
and two decay constants ($f_q$ and $f_s$), i.e.~\cite{FKS},
 \be\label{fqfs}
 \left( \begin{array}{cc}
 f^q_\eta  \;\;\;\;\; f^s_\eta\\
 f^q_{\eta'} \;\;\;\;\; f^s_{\eta'}
 \end{array}\,\right)
 = \left( \begin{array}{cc}
 \cos\phi_q\;\; -\sin\phi_s\\
 \sin\phi_q\;\;\;\;\;\cos\phi_s
 \end{array}\,\right)\left( \begin{array}{cc}
 f_q\;\; 0\\
 0\;\;\;f_s
 \end{array}\,\right).
 \ee
The difference between the mixing angles $\phi_q-\phi_s$ is due to the
Okubo-Zweig-Iizuka(OZI)-violating effects~\cite{OZI} and is found to be
small ($\phi_q-\phi_s<5^{\circ}$).
The OZI rule implies that the difference between
$\phi_q$ and $\phi_s$ vanishes (i.e., $\phi_q=\phi_s=\phi$)
to leading order in the $1/N_c$ expansion. Similarly, the four parameters
$f^{8}_P$ and $f^{0}_P$ in the octet-singlet basis
may be written in terms of two angles ($\theta_8$ and $\theta_0$)
and two decay constants ($f_8$ and $f_0$). However, in this
case, $\theta_8$ and $\theta_0$ turn out to differ
considerably and become equal only in the ${\rm SU}_f(3)$ symmetry limit~\cite{FKS,Leut98}.

\end{document}